\documentclass[usegraphicx,useAMS,usenatbib]{mn2e}
\usepackage{graphicx}

\newcommand{\lii}{Li\,{\footnotesize I}}

\newcommand{\kms}{\,km\,s$^{-1}$}

\newcommand{\be}{\begin{equation}}
\newcommand{\ee}{\end{equation}}
\newcommand{\bd}{\begin{displaymath}}
\newcommand{\ed}{\end{displaymath}}

\title[The Gamma Vel association]
  {The stellar association around Gamma Velorum and its relationship
  with Vela OB2}
\author[R.D. Jeffries et al.]
  {R.D.~Jeffries$^1$,  Tim Naylor$^2$, F.M. Walter$^3$, M.P. Pozzo$^4$
   and C.R. Devey$^1$\\
  $^1$ Astrophysics Group, Research Institute for the Environment, Physical
  Sciences and Applied Mathematics, Keele University, \\ Keele, 
      Staffordshire ST5 5BG\\
  $^2$ School of Physics, University of Exeter, Stocker Road, Exeter
  EX4 4QL\\
  $^3$ Department of Physics and Astronomy, State University of New York, Stony Brook, NY 11794-3800, USA\\
  $^4$ Department of Earth Sciences and Materials Simulation
  Laboratory, University College London, Gower Street, London WC1E 6BT\\ 
}
\date{Submitted September 2008}

\pagerange{\pageref{firstpage}--\pageref{lastpage}} \pubyear{2007}

\def\LaTeX{L\kern-.36em\raise.3ex\hbox{a}\kern-.15em
    T\kern-.1667em\lower.7ex\hbox{E}\kern-.125emX}

\begin{document}

\label{firstpage}

\maketitle

\begin{abstract}
We present the results of a photometric $BVI$ survey of 0.9 square
degrees around the Wolf-Rayet binary $\gamma^2$~Vel and its early-type
common proper motion companion $\gamma^1$~Vel (together referred to as the
$\gamma$~Vel system). Several hundred pre-main-sequence (PMS) stars are
identified and the youth of a subset of these is spectroscopically
confirmed by the presence of lithium in their atmospheres, H$\alpha$
emission and high levels of X-ray activity. We show that the PMS stars
are kinematically coherent and spatially concentrated around
$\gamma$~Vel. The PMS stars have similar proper motions to
$\gamma$~Vel, to main-sequence stars around $\gamma$~Vel and to
early-type stars of the wider Vela OB2 association of which
$\gamma^2$~Vel is the brightest member. The ratio of main-sequence
stars to low-mass (0.1--0.6\,$M_{\odot}$) PMS stars is consistent with
a Kroupa (2001) mass function. Main-sequence fitting to stars around
$\gamma$~Vel gives an association distance modulus of $7.76\pm 0.07$\,
mag, which is consistent with a similarly-determined distance for Vela
OB2 and also with interferometric distances to
$\gamma^2$~Vel. High-mass stellar models indicate an age of 3--4\,Myr
for $\gamma^2$~Vel, but the low-mass PMS stars have ages of $\simeq
10$\,Myr according to low-mass evolutionary models and 5--10\,Myr by
empirically placing them in an age sequence with other clusters based
on colour-magnitude diagrams and lithium depletion.  We conclude that
the low-mass PMS stars form a genuine association with $\gamma$~Vel and
that this is a subcluster within the larger Vela OB2 association. We
speculate that $\gamma^2$~Vel formed after the bulk of the low-mass
stars, expelling gas, terminating star formation and unbinding the
association. The velocity dispersion of the PMS stars is too low for
this star forming event to have produced all the stars in the extended
Vela OB2 association. Instead, star formation must have been initiated
at several sites within a molecular cloud, either sequentially or,
simultaneously after some triggering event.
\end{abstract}


\section{Introduction}

The double lined spectroscopic binary system $\gamma^2$ Vel (HD 68273)
contains the closest example of a Wolf-Rayet star. The two components
of the binary orbit with a period of 78.5 days and have spectral types
WC8 (Smith 1968) and O8III (Schaerer, Schmutz \& Grenon 1997).
Although the Hipparcos parallax to the system yields a distance of only 
$258^{+41}_{-31}$\,pc, de Zeeuw et al. (1999) classify it as belonging to the Vela
OB2 association, a group of some 100 early-type stars spread over an
angular diameter of $\sim 10^{\circ}$ and at a mean distance of $410\pm
12$\,pc. Indeed, the distance to $\gamma^2$ Vel is somewhat
controversial. Millour et al. (2007) found an interferometric distance
of $368^{+38}_{-13}$\,pc, while North et al. (2007) have combined
interferometry with radial velocity measurements to estimate a
distance of $336^{+8}_{-7}$\,pc. A revised analysis of the Hipparcos
data by van Leeuwen (2007) quotes a distance of
$334^{+40}_{-32}$\,pc. A further common proper-motion component of the
system, $\gamma^1$ Vel (HD 68243), lies 41 arcseconds to the south-west  
of $\gamma^2$ Vel. It is an SB1 binary with a B2III primary and a
period of 1.48~days (Hern\'andez \& Sahade 1980). The system formed by
$\gamma^1$ and $\gamma^2$~Vel will hereafter be referred to as $\gamma$~Vel.

Pozzo et al. (2000) presented results of {\it ROSAT} X-ray observations
combined with a photometric survey in a region surrounding $\gamma$
Vel. A large number of X-ray sources were found
coinciding with low-mass stars that were identified with a few Myr
pre-main-sequence (PMS) isochrone at a distance of 350--450\,pc. Pozzo et
al. argued that these low-mass stars were physically associated with
$\gamma$~Vel and that they are all part of the Vela OB2 association.

In this paper we present a $BVI$ photometric survey of about 0.9 square
degrees around $\gamma$ Vel which is complete to $V\sim 20$. This area
is more extended than that considered by Pozzo et al. (2000) and allows
us to revisit the question of the spatial distribution of the low-mass
association. We present optical spectroscopy of a subsample of the PMS
candidates which confirms both the youth and kinematic coherence of the
association. We have also used {\it XMM-Newton} observations around
$\gamma$ Vel to examine the X-ray properties of the low-mass PMS
candidates and define a secure sample of association members with which
to investigate their age. Finally, by combining what we have
learned about the low-mass PMS stars with distances obtained from
modelling upper main sequence stars around $\gamma$~Vel and the age of
$\gamma$~Vel itself, we examine the relationship between the high- and
low-mass members of this group and its place within the wider Vela OB2
association.

\section{Observational Data}

\subsection{Optical photometry}
\label{photometry}

\begin{table}
\caption{A log of the observations used in this work. For each of the
  fields below, we obtained exposures of (300, 20s), (120, 10s) and
  (60, 6s) through the $B$, $V$ and $I$ filters respectively.}
\begin{tabular}{lccc}
\hline
Field & Date & \multicolumn{2}{c}{Field Centre (J2000)}  \\
      &      &  RA   &   Dec  \\
\hline
01 & 1999-02-09 & 08 10 17& -47 20 29  \\  
02 & 1999-02-09 & 08 09 36& -47 13 32  \\ 
03 & 1999-02-09 & 08 08 52& -47 20 39  \\  
04 & 1999-02-09 & 08 09 32& -47 27 48  \\ 
05 & 1999-02-09 & 08 10 35& -47 09 47  \\ 
06 & 1999-02-09 & 08 08 24& -47 09 44  \\  
07 & 1999-02-09 & 08 08 24& -47 31 42  \\  
08 & 1999-02-09 & 08 10 34& -47 31 38  \\ 
09 & 2002-02-09 & 08 09 31& -47 09 24  \\  
10 & 2002-02-09 & 08 10 26& -47 20 23  \\ 
11 & 2002-02-09 & 08 09 31& -47 31 24  \\ 
12 & 2002-02-09 & 08 08 26& -47 20 23  \\ 
13 & 2002-02-12 & 08 07 22& -47 41 42  \\ 
14 & 2002-02-12 & 08 07 22& -47 31 22  \\ 
15 & 2002-02-11 & 08 07 22& -47 20 23  \\ 
16 & 2002-02-11 & 08 07 22& -47 09 22  \\ 
17 & 2002-02-11 & 08 07 22& -46 58 23  \\ 
18 & 2002-02-11 & 08 08 26& -46 58 21  \\ 
19 & 2002-02-11 & 08 09 31& -46 58 22  \\ 
20 & 2002-02-11 & 08 10 26& -46 58 23  \\ 
21 & 2002-02-11 & 08 11 30& -46 58 23  \\ 
22 & 2002-02-11 & 08 11 30& -47 09 23  \\ 
23 & 2002-02-11 & 08 11 30& -47 20 23  \\ 
24 & 2002-02-12 & 08 11 30& -47 31 24  \\ 
25 & 2002-02-12 & 08 11 30& -47 41 24  \\ 
26 & 2002-02-12 & 08 10 26& -47 41 22  \\ 
27 & 2002-02-12 & 08 09 31& -47 41 23  \\  
28 & 2002-02-12 & 08 08 26& -47 41 22  \\ 
\hline
\end{tabular}
\label{photom}
\end{table}

A photometric survey of an area around $\gamma$ Vel was performed at
the Cerro Tololo Interamerican Observatory (CTIO) 0.9-m telescope. The survey
was done in two parts in different years, but using the same
instrumentation and filters, namely a Tek 2048$\times$2048 CCD with a
$13.5\times13.5$ arcmin$^2$ field of view, the Harris $B$ and $V$
filters and a Kron-Cousins $I$ filter.

The first set of observations, described in Pozzo et al. (2000),
consisted of eight overlapping fields surveyed on the night beginning 8 February 1999,
together with five standard star fields taken from the Landolt (1992)
catalogue. Both short (20, 10, 6s) and long (300, 120, 60s) exposures
were taken through the $B,V,I$ filters at each of the field centres listed in
Table~\ref{photom}. The data were taken in photometric conditions.\footnote{ 
The exposure times of (20, 10, 10s) and (200, 100, 100s) 
previously stated by Pozzo et al. (2000) were incorrect, but this error
was not present in the calculated magnitudes in that paper.}

The second set of observations consisted of twenty fields which were
surveyed on the nights beginning 8, 10 and 11 February 2002. These twenty fields had
substantial overlap with the original eight fields and the field
centres are given in Table~\ref{photom}. The short and long
exposures also consisted of (20, 10, 6s) and (300, 120, 60s) through the
$B,V,I$ filters respectively. The data were obtained in good to
photometric conditions and again observations of several Landolt (1992) fields
were taken each night, some on more than one occasion.
The combined 28 overlapping fields make a survey of roughly constant
sensitivity over a rectangular area of approximately
$0.93\times0.95$ degrees, centred upon $\gamma^2$ Vel.

The data were initially reduced by subtracting a median stacked bias
frame and overscan and then normalising with flat fields taken during
twilight sky on each night. A bad pixel mask was constructed to
identify hot pixels and a few bad columns on the CCD.  The data were
analysed using the {\sc cluster} software package described by Naylor
et al. (2002), with updates described in Burningham et al. (2003) and
Jeffries et al. (2004). Objects were identified in the long $V$-band
exposures (or short $V$ exposures if saturated in the long). Spatial
transformations were calculated between the separate exposures for each
field and optimal photometry (Naylor 1998) was performed to obtain a
flux for each object in each exposure at a given position.  A spatially
varying ``profile correction'' was calculated from bright, unsaturated
stars in each frame so that the photometry could later be calibrated
with standard star observations. The profile-corrected photometry in
each band was combined in a weighted fashion after adjustments were
made for any airmass difference. An additional statistical uncertainty
was added to each measurement at this stage to ensure that a plot of
signal-to-noise versus chi-squared for the combined measurements was
flat and approximately equal to unity (see Naylor et al. 2002 for
details). This additional uncertainty ranged from 0.007 to 0.015 mag
for the 28 fields and reflects uncertainties in the profile correction.
Astrometric calibration of each frame was performed using the positions
of 600--900 stars from the 2MASS point source catalogue (Cutri et
al. 2003). The rms residuals were approximately 0.08 arcseconds in each
coordinate.

Standard star magnitudes were measured through a 15 pixel (6
arcsec) radius aperture and weighted least squares solutions to the
following equations were found for each night.
\begin{eqnarray}
B-V & = & \psi_{bv}\, (b - v) - k_{bv}X + z_{bv}\, ,\\
V-I & = & \psi_{vi}\, (v - i) - k_{vi}X + z_{vi}\, ,\\
V   & = & v + \psi_v\, (B-V) - k_v X + z_v\, ,
\end{eqnarray}
where $b,v,i$ are the instrumental magnitudes, $X$ is the airmass, $k$ are the extinction
coefficients, $z$ the zeropoints and $\psi$ are the colour terms.
Standard stars were measured with a range of colours, but included only two
with $V-I>2.5$ (the reddest had $V-I=5.8$).  
The means of the colour terms were found to be
$\psi_{bv}=0.89$, $\psi_{vi}=1.00$ and $\psi_{v}=0.02$, indicating that
the filter set used was close to the natural Johnson-Kron-Cousins
system defined by the standards.  Using these transformations the
magnitudes were place onto the Johnson-Kron-Cousins $BVI$ system.  The
last step in the construction of the catalogue was to make the
photometric calibration uniform across the mosaic using stars in the
overlapping regions of each frame (see Jeffries et al. 2004). The
adjustments made here were generally less than 0.02 mag and the
remaining rms differences for stars in the overlap regions was about
0.01 mag in $V$, $B-V$ and $V-I$.

\begin{table*}
\caption{The photometric catalogue for a 0.9 square degree area around
  $\gamma$ Vel. Only a few rows are shown here to illustrate the
  content, the full table is available in electronic form. The columns
  list two identifying numbers (the first of which is the field number
  from Table~\ref{photom}), the RA and Dec (J2000), an x,y pixel position on
  the CCD where the object was identified and then for each of $V$,
  $B-V$ and $V-I$ there is a magnitude, uncertainty and a quality flag. The
  flagging system is described in detail by Burningham et al. (2003).}
\begin{tabular}{rrc@{\hspace*{1mm}}c@{\hspace*{1mm}}cc@{\hspace*{1mm}}c@{\hspace*{1mm}}cc@{\hspace*{1mm}}cc@{\hspace*{2mm}}c@{\hspace*{2mm}}cc@{\hspace*{2mm}}c@{\hspace*{2mm}}cc@{\hspace*{2mm}}c@{\hspace*{2mm}}c}
\hline
\multicolumn{2}{c}{Identifier} & \multicolumn{3}{c}{RA} &
\multicolumn{3}{c}{Dec} &  x & y &\multicolumn{3}{c}{$V$} &
\multicolumn{3}{c}{$B-V$} & \multicolumn{3}{c}{$V-I$} \\
\hline
  14 &   6058&  08& 06& 38.742& -47& 37& 17.96&  2057.835&  1900.018&     21.483&      0.680&  EN&    2.936&      0.898&  EN&    1.789&      0.868&  NN\\
  14 &    413&  08& 06& 38.756& -47& 33& 56.51&  2058.757&  1398.062&     17.033&      0.029&  EO&    6.987&      0.667&  EO&    0.878&      0.028&  OO\\
  14 &   1571&  08& 06& 38.756& -47& 34& 46.17&  2058.448&  1521.786&     18.915&      0.025&  EO&    5.285&      0.667&  EO&    1.883&      0.070&  OO\\
  14 &   4261&  08& 06& 38.787& -47& 36& 03.27&  2057.166&  1713.902&     20.921&      0.098&  EO&    3.462&      0.672&  EO&    1.337&      0.159&  OO\\
  14 &   1040&  08& 06& 38.838& -47& 35& 43.23&  2056.024&  1663.976&     18.323&      0.016&  OO&    0.723&      0.021&  OO&    0.889&      0.029&  OO\\
\multicolumn{19}{c}{...} \\
\hline
\end{tabular}
\label{catalog}
\end{table*}

\begin{figure*}
\centering
\begin{minipage}[t]{0.45\textwidth}
\includegraphics[width=80mm]{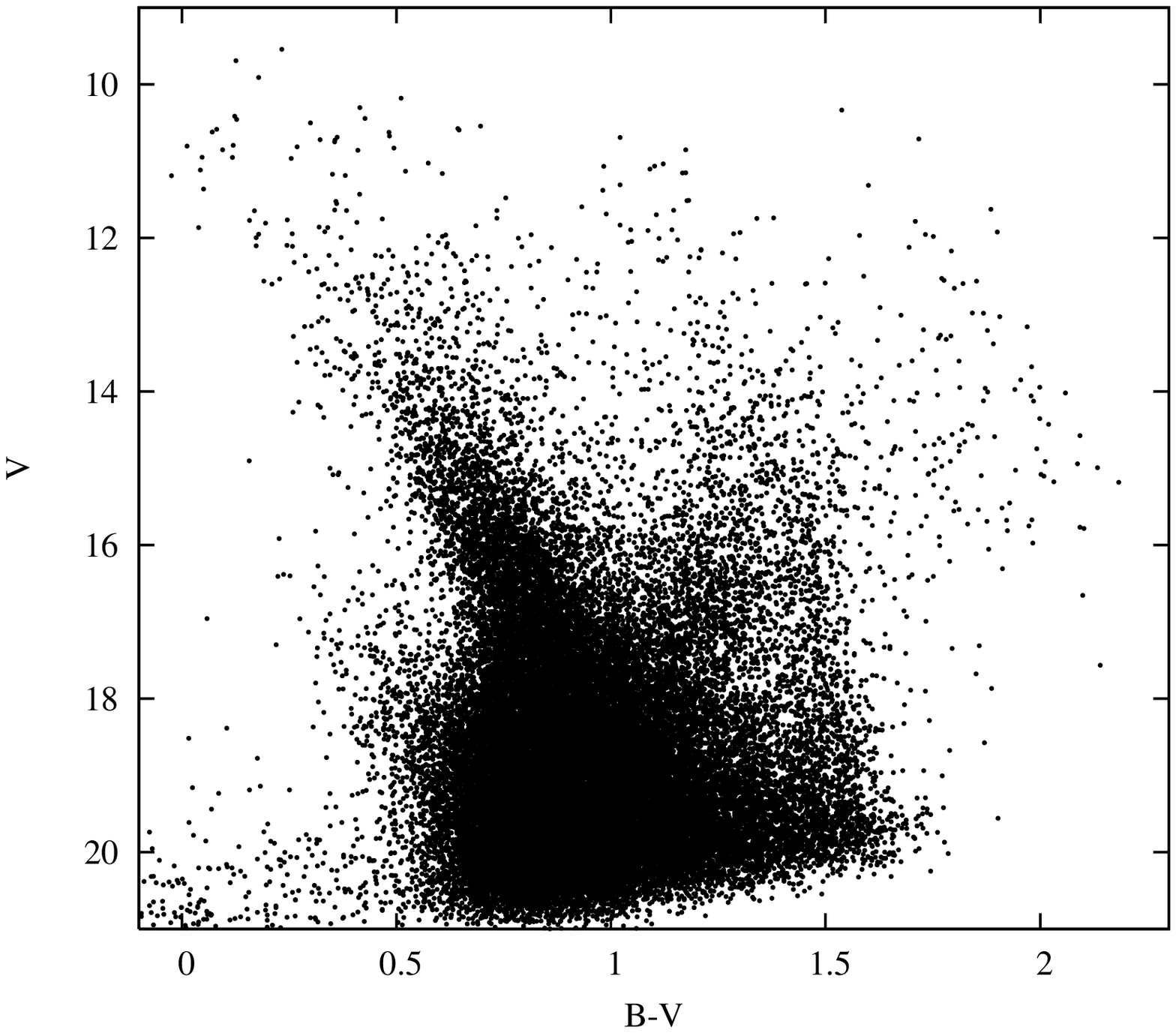}
\end{minipage}
\begin{minipage}[t]{0.45\textwidth}
\includegraphics[width=80mm]{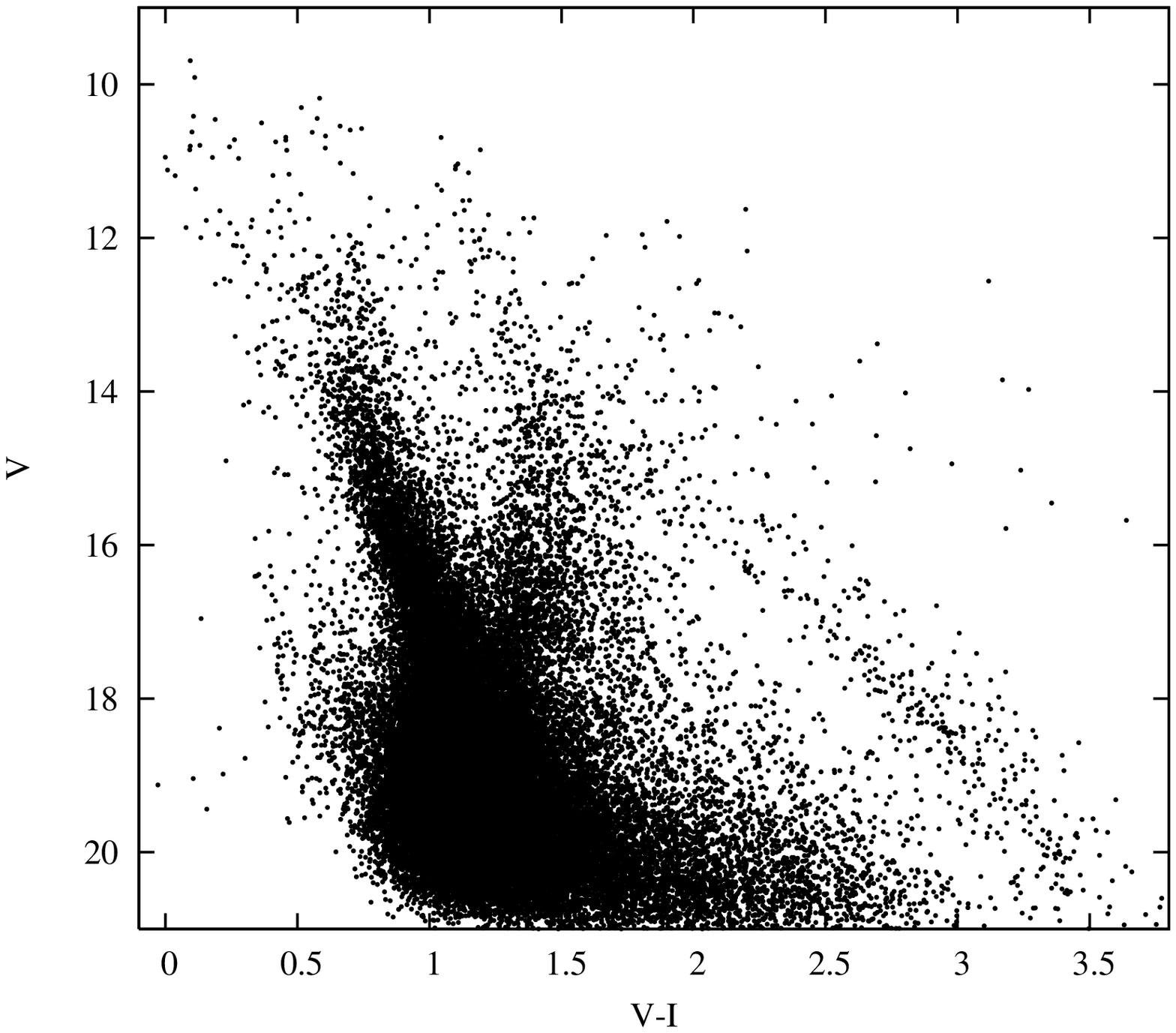}
\end{minipage}
\caption{Colour-magnitude diagrams (CMDs) from the entire survey
  area. Only unflagged photometric points with uncertainties less than
  0.1 mag in each coordinate are plotted -- 38345 stars for the $V$,
  $B-V$ diagram and 40185 stars for the $V$, $V-I$ diagram. A clear PMS
  stands out in the $V$ versus $V-I$ CMD for $V-I>1.6$.
}
\label{cmd}
\end{figure*}

The final optical catalogue is presented in Table~\ref{catalog}
(available fully in electronic form only). The flags applied to the
photometric values (non-stellar, bad sky, saturation etc.) are detailed
in Burningham et al. (2003). Some of these flags, particularly the
ill-determined background (flagged ``I''), are quite conservative but
this ensures that the unflagged stars have photometry of very high quality.
The final photometric uncertainties listed in the catalogue include the
additional systematic error associated with the profile correction but
do not include the 0.01 mag rms found for stars in the overlapping
field regions. This additional error should be added when comparing or
combining stars across different fields or the whole survey (as in this
paper). Finally there are likely to be external systematic photometric
uncertainties associated with placing our photometry onto the standard
system. The residuals of the fits to the standard star magnitudes
suggest that any uncertainties are limited to less than 0.02 mag in the
reported magnitudes and colours for stars with $V-I\leq 2.5$. This may
rise for redder stars where there is a paucity of available photometric
standards.

The final colour-magnitude diagrams (CMDs) are shown in Fig.~\ref{cmd}
for unflagged stars with statistical uncertainties of less than 0.1 mag
on each axis. The PMS identified by Pozzo et al. (2000) is readily
appearent, especially in the $V$ vs $V-I$ CMD where for $V-I>1.6$ it
stands well clear of contamination. In the $V$ vs
$B-V$ CMD there appear to be some very faint blue objects. Upon visual
inspection most of these appear to be situated along the diffraction
spikes of bright, presumably early-type stars, but the detections were
too faint for them to be correctly identified as non-stellar.

A histogram of $V$ magnitudes suggests that the $V$ vs $V-I$ CMD is
almost complete to $V=20$ for colours and magnitudes with uncertainties
$<0.1$ mag.  The catalogue is ``almost'' complete to this level because
there are instances where brighter stars are flagged as non-stellar or
have a bad or ill-determined sky flag. This is especially true in a 3
arcminute radius immediately around $\gamma$ Vel itself, where almost
every star is saturated or has photometry affected by scattered
light from $\gamma$ Vel, but is also the case for much smaller regions
around other bright stars in the survey area. The completeness at the
bright end is governed by the saturation level in the short exposures
which is at about $V=10.5$ mag, with small variations across the mosaic.
Excluding the central 3 arcminute radius, 86 per cent of stars with
$11<V<20$ have unflagged photometry and photometric uncertainties of
$<0.1$ mag. This fraction is not very
magnitude dependent, falling from $\simeq 93$ per cent for $V<16$ to
83 per cent for $19<V<20$.

\subsection{Optical Spectroscopy}

\begin{figure*}
\centering
\begin{minipage}[t]{0.45\textwidth}
\includegraphics[width=80mm]{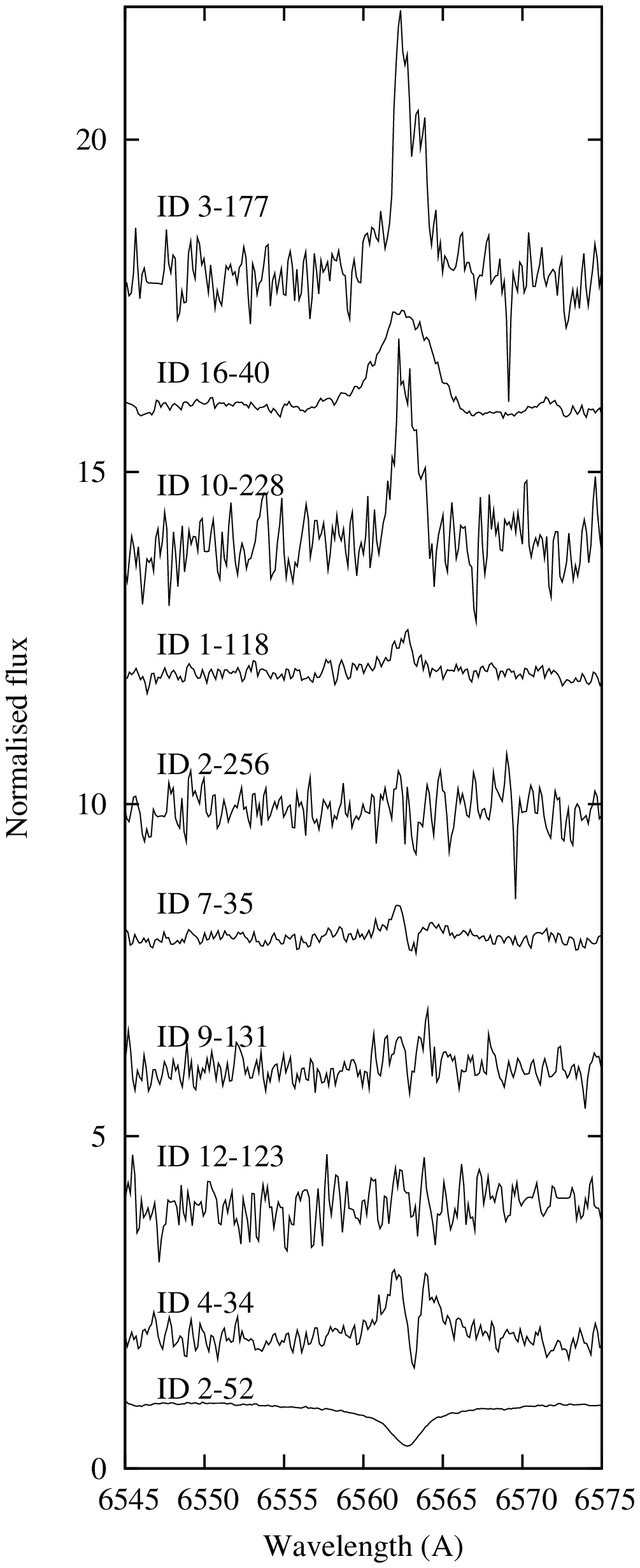}
\end{minipage}
\begin{minipage}[t]{0.45\textwidth}
\includegraphics[width=80mm]{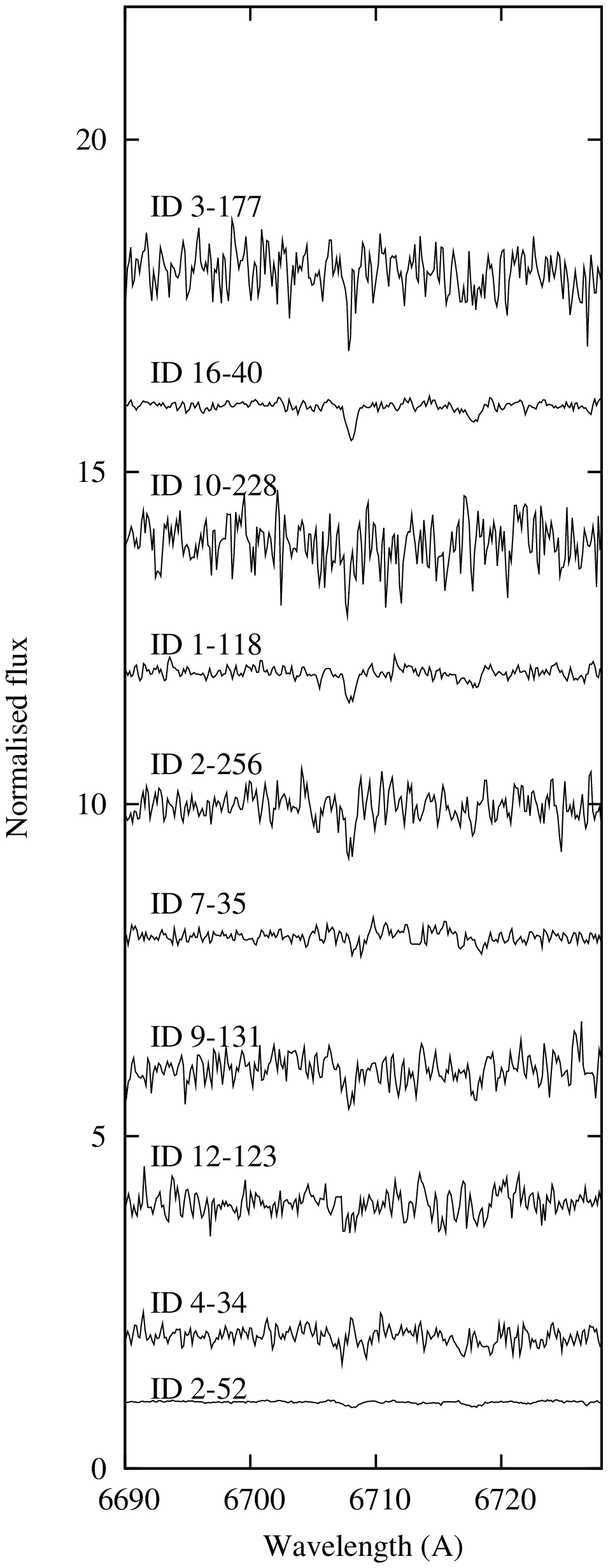}
\end{minipage}
\caption{Examples of our spectra in the vicinities of the
  H$\alpha$ and Li\,{\sc i} 6708\AA\ lines. The spectra have been
  normalised to unity and offset for clarity. These examples show the
  full diversity of signal-to-noise ratios in our sample.
}
\label{specplot}
\end{figure*}

\begin{table*}
\caption{Spectroscopy results. Only a few rows are shown here to
  illustrate the content. The full table is available in electronic
  form. Column 1 lists which night the spectra were obtained: $1=$ 24
  November 1999, $2=$ 25 November 1999. Columns 2 and 3 give the source
  identification number (as in Table~2).  Columns 4--14 (omitted from
  the table sample shown below, but included for convenience in the
  electronic version) repeats the Table~2 values of RA, Dec (J2000) and
  the photometry values, uncertainties and flags for these
  sources. Column 15 gives the signal-to-noise ratio per pixel for the
  spectrum estimated from the continuum around the Li\,{\sc i}
  line. Columns 16 and 17 give the EW and EW uncertainty for the
  Li\,{\sc i} 6708\AA\ line -- a zero uncertainty indicates that the
  quoted W[Li] is a 2-sigma upper limit. Columns 18 and 19 give the
  emission EW of the H$\alpha$ line -- here a negative value indicates
  absorption. Column 20 gives the heliocentric RV measured from the
  spectrum. Columns 21 to 24 list the Nomad proper motions and
  uncertainties (in milli-arcsec/year) where available.
  Column 25 gives a comment based on the cross-correlation
  function (ccf) used to derive the RV: SB2 indicates a double peaked
  ccf, SB2? indicates a possible double peaked ccf, RR means the star
  is probably a rapid rotator (in excess of $\simeq 30$\,km\,s$^{-1}$)
  and E indicates an early-type spectrum. Column 26 lists $\log (L_{\rm
  x}/L_{\rm bol})$ where available (see section~\ref{xraymembers}).
  Column 27 gives our
  assessment of association membership: 1 indicates a member, 2
  indicates a non-member and 3 indicates uncertain membership (see
  section~\ref{specmembers}).}
\begin{tabular}{r@{\hspace*{2mm}}r@{\hspace*{2mm}}r@{\hspace*{2mm}}c@{\hspace*{2mm}}c@{\hspace*{2mm}}c@{\hspace*{2mm}}c@{\hspace*{2mm}}c@{\hspace*{2mm}}c@{\hspace*{2mm}}c@{\hspace*{2mm}}c@{\hspace*{2mm}}c@{\hspace*{2mm}}c@{\hspace*{2mm}}c@{\hspace*{2mm}}c@{\hspace*{2mm}}c}
\hline
 (1)  & (2) & (3) & (15) & (16) &(17) &(18)&(19)&(20)&(21)&(22)&(23)&(24)&(25)&(26)&(27)\\
Night & \multicolumn{2}{c}{Identifier} & SN & W[H$\alpha$] &
$\delta$W[H$\alpha$] & W[Li] & $\delta$W[Li] & RV & PMRa & $\delta$PMRa
& PMDec & $\delta$PMDec&Comment & $\log$ &Member \\
      &                                &    &  & (\AA)       &
(\AA)                & (\AA) & (\AA)         & (km\,s$^{-1}$) & \multicolumn{4}{c}{(mas/yr)}& &$(L_{\rm x}/L_{\rm bol})$&\\
\hline
1  &  16 &  40 &   22 &  4.99& 0.29 & 0.49 &0.07 &   18&-5.7&5.6&13.4&5.6&  &    &  1\\
1  &   7 &  35 &   14 & -0.40& 0.31 & 0.31 &0.08 &   21&0.6&5.6&12.4&5.6&   &-2.91   &  1\\
1  &   6 & 216 &    5 &  4.3 & 2.2  & 0.43 &0.15 &   14&   &   &    &  & & & 1\\
1  &   9 & 131 &    6 &  0.52& 0.56 & 0.50 &0.13 &   20&-5.2&5.6&2.4&5.6&   &-3.45   &  1\\
2  &  12 & 123 &    6 &  0.21& 0.26 & 0.46 &0.13 &   30&-11.1&5.7&4.9&5.6& SB2? &-3.82&  1\\
\multicolumn{16}{c}{...}\\
\hline
\end{tabular}
\label{specresults}
\end{table*}

On 24 and 25 November 1999 we used the 4-m Blanco telescope at CTIO in
conjunction with the Hydra multi-fibre spectrograph to obtain high
resolution ($R\simeq  25000$) spectra of 120 candidate members of the
low-mass association around $\gamma$ Vel. Targets were selected
from the original photometric survey discussed in Pozzo et al. (2000)
(the eight fields observed in February 1999) and so come from the central
$\sim 0.3$ square degrees of our total optical survey. Targets were
chosen solely on the basis of their position in the $V$ vs $V-I$ CMD.

The 31.6 l/mm echelle grating and filter \#5 were used to
isolate order 8, achieving a wavelength coverage from about 6450\AA\ to 6750\AA.
The detector was a Loral 3k with 15 micron pixels. We operated at gain setting
4, which yields about 2 electrons per ADU and 7.5 electrons of read noise
per pixel.

The same field centre was observed twice on successive nights, observing
mainly different stars on each occasion.  There were 70 fibres
in use at the time of our observations.  We observed 62 and 58 targets
in the two observations with $3\times 600$s exposures. A few targets
were included in both configurations, so a total of 112 separate
targets were observed.
Sky spectra were obtained through 7 and 6
fibres, respectively, on the two nights. Not all fibres were used
because of targeting constraints.  Spectra of the daylight sky
were used to establish fibre throughputs - with two exceptions they vary by less
than a factor of 2, although two fibers had a very low throughput.  
One or two stars were abnormally faint (for their magnitude) in each field.  We
attribute this to mispositioning of the fibre. 

A Th-Ar calibration lamp was observed at the start and end of the night to
establish the dispersion solution. An etalon lamp was observed before and
after each observation to track shifts in the zero-point of the wavelength
solution. The data were flattened using dome flats illuminated by a quartz
lamp.

Data reductions and extractions were accomplished using standard {\sc
iraf} tasks including the extraction of 1-dimensional spectra and
application of the dispersion solution. A scattered light correction
was applied.  The twilight sky was observed each night to establish the
zero point of the radial velocities. Some examples of the spectra are
shown in Fig.~\ref{specplot}.

Radial velocities (RVs) were measured by
cross-correlating the spectra against those of the sky taken through
the same fibre. The statistical uncertainties are estimated to be
about 2--3\kms\ for the reported RVs, although are probably
worse for a few objects with the lowest signal-to-noise. In some cases
no meaningful cross-correlation peak could be measured.

The equivalent widths (EWs) of the \lii~6708\AA\ line (W[Li]) were
estimated using the {\sc splot} task in {\sc iraf} and a continuum
estimated from a clipped fit to the surrounding continuum. Estimated
uncertainties are based upon the formula $\delta$W[Li]$=\sqrt{(fp)/SN}$
where $f$ and $p$ are the full width half maximum of the line (about
0.7\AA\ for most objects) and the pixel size (about 0.14\AA), and $SN$
is the signal-to-noise ratio of the spectrum which is empirically
estimated from the residuals to the continuum fit described
above. Where the line was not significantly detected, a 2-sigma upper
limit is reported.

H$\alpha$ EWs were estimated in a similar way. However it should be
noted that the strong H$\alpha$ background signal from the sky, as seen
in the sky fibres, meant that the signal-to-noise ratio in the
H$\alpha$ line was usually much worse than the surrounding
continuum. The EW uncertainties are dominated by errors in the fibre
throughput estimates in the fainter stars. The contribution of this to
the EW uncertainties was estimated using the standard deviation of the
residual flux around H$\alpha$ seen in the sky-subtracted spectra from
fibres placed on the night sky.

The results of these analyses are presented in Table~\ref{specresults}.
In many cases, especially for spectra from the second night or for
faint targets, the spectra were too poor to provide any reliable
information and so Table~\ref{specresults} only contains results from
91 spectra of 84 individual targets. Note that seven targets have
photometry flags set -- indicating that their photometry cannot be
considered as reliable as for the rest. The initial photometry analysis
of Pozzo et al. (2000) did not flag these stars, indicating that the
more stringent quality control applied here, particularly with regard
to``ill-determined sky'' (see section~\ref{photometry}), has flagged
these stars but that their photometry cannot be of very poor
quality. Five of these stars are judged to be members in
section~\ref{specmembers} and their photometry places them squarely in
the cluster pre-main-sequence. We will continue to use these stars in
our analysis and in any event, their inclusion affects none of the
paper's conclusions.

\subsection{XMM-Newton Observations}

\label{xmm}

The X-ray data used in this paper come from two observations performed
in 2001 by the {\it XMM-Newton} satellite. The first started at at MJD 52013.473 and
lasted 29.7\,ks and the second began at MJD 52033.4213 and
lasted 59.1\,ks. The central pointing positions were close, but slightly different
for each observation at RA\,$=122.407208$ degrees, Dec\,$=-47.358139$ degrees
and RA\,$=122.397042$, Dec\,$=-47.362889$ degrees respectively. Both
observations made use of the European Photon Imaging Camera (EPIC) 
camera with the ``thick'' optical filter covering the pn, m1 and m2
detectors (see Str\"uder et al. 2001; Turner et al. 2001).

These EPIC datasets were reduced as part of the second XMM-Newton
Serendipitous Source Survey (2XMM) cataloguing effort and the
positions and X-ray count rates for sources detected in these two
observations were obtained from the 2XMM database (pre-release version) held at
the XMM-Newton Survey Science Centre at Leicester University.

The data available consist of the positions and detection likelihoods,
together with count rates determined in a number of separate energy
bands and their corresponding uncertainties. For the purposes of this
paper we created a merged X-ray catalogue consisting of all those
sources in the longer observation plus those sources in the shorter
observation with no counterpart (within 6 arcseconds) in the longer
observation. Anything correlated within 6 arcseconds was assumed to be
the same X-ray source -- there is only a likelihood of 1 random
correlation within this radius.  The merged X-ray catalogue contained
positions and count rates for 276 separate sources, of which just 26 were
found only in the shorter observation.

\section{Association Membership}

\label{members}

\subsection{Spectroscopic members}

\label{specmembers}

\begin{figure*}
\centering
\begin{minipage}[t]{0.45\textwidth}
\includegraphics[width=80mm]{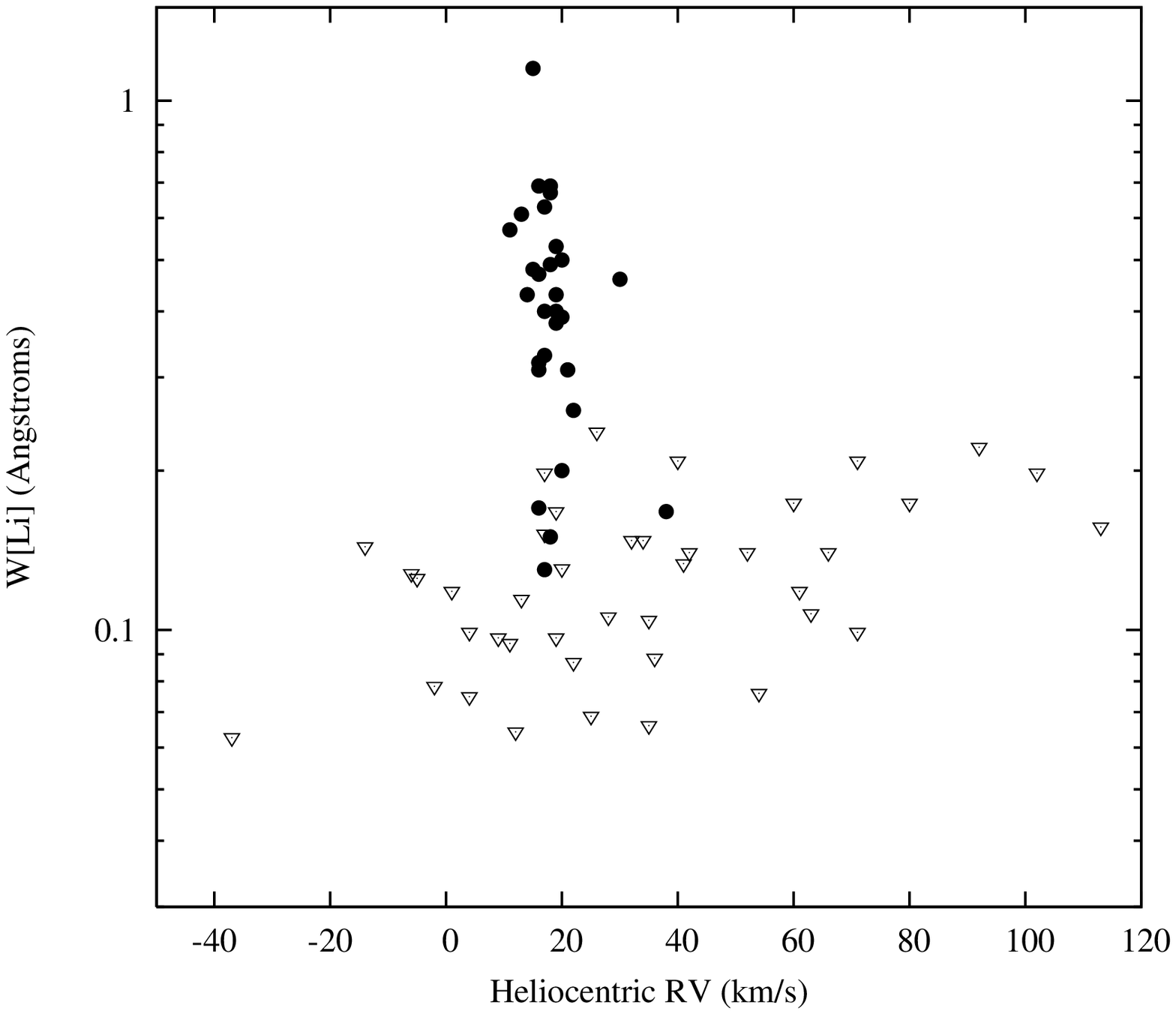}
\end{minipage}
\begin{minipage}[t]{0.45\textwidth}
\includegraphics[width=80mm]{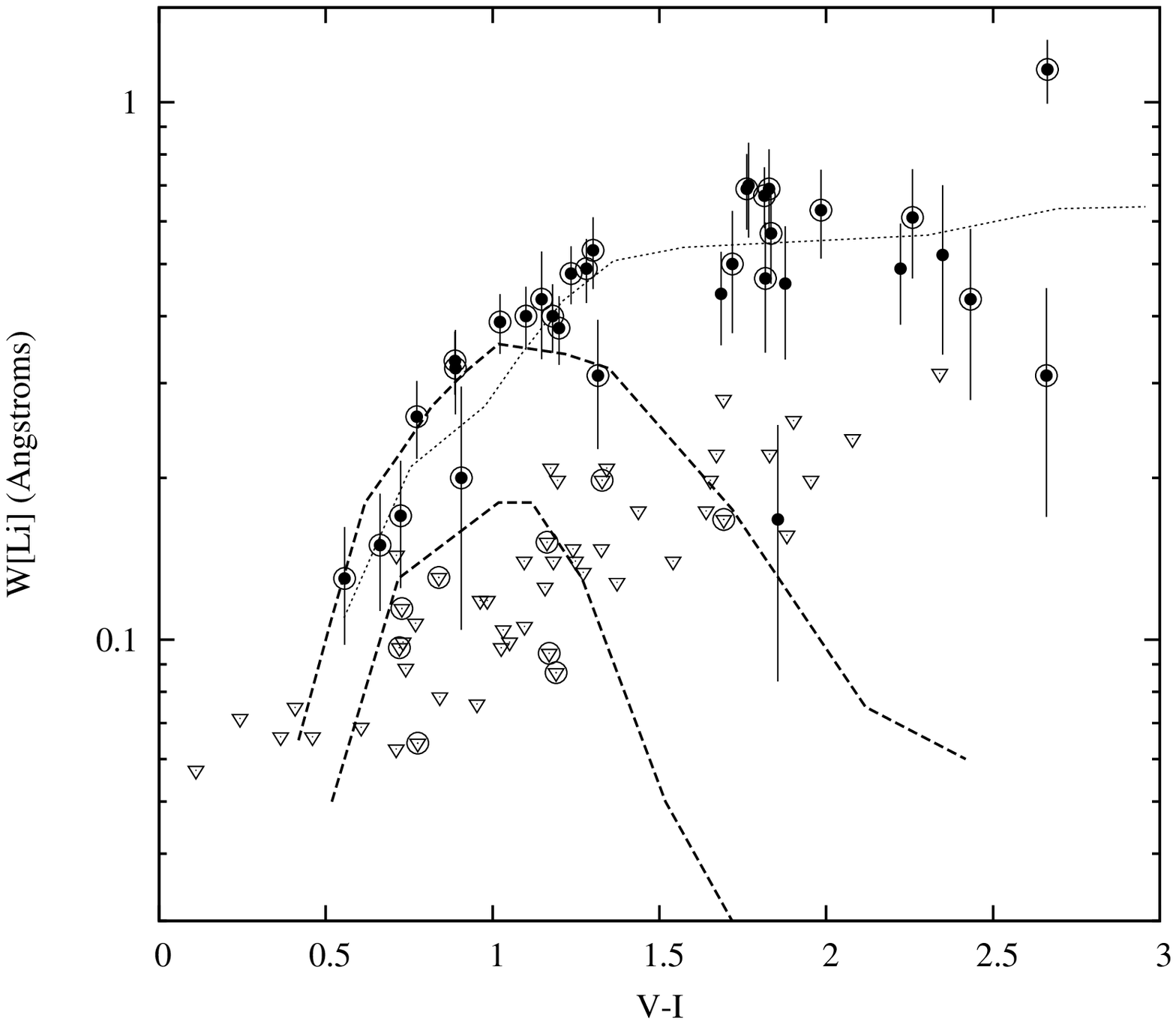}
\end{minipage}
\caption{(a) Strength of the Li\,{\sc i}~6708\AA\ feature versus the
  heliocentric RV. The triangles represent upper limits.
  There is a clear concentration of objects exhibiting
  strong Li in a narrow range of RV. (b) Strength of the Li\,{\sc i}~6708\AA\
  feature versus $V-I$ colour. Encircled objects have $11\leq$RV$\leq
  22$\,km\,s$^{-1}$. The dashed loci show observationally determined ranges
  of W[Li] in the 30--50\,Myr old open clusters IC~2391, IC~2602
  and NGC~2547 open clusters. The dotted locus corresponds to a Li
  abundance $A$(Li)$=3.1$ which approximates a star that has not
  depleted its initial Li.}
\label{lirv}
\end{figure*}

\begin{figure*}
\centering
\begin{minipage}[t]{0.45\textwidth}
\includegraphics[width=80mm]{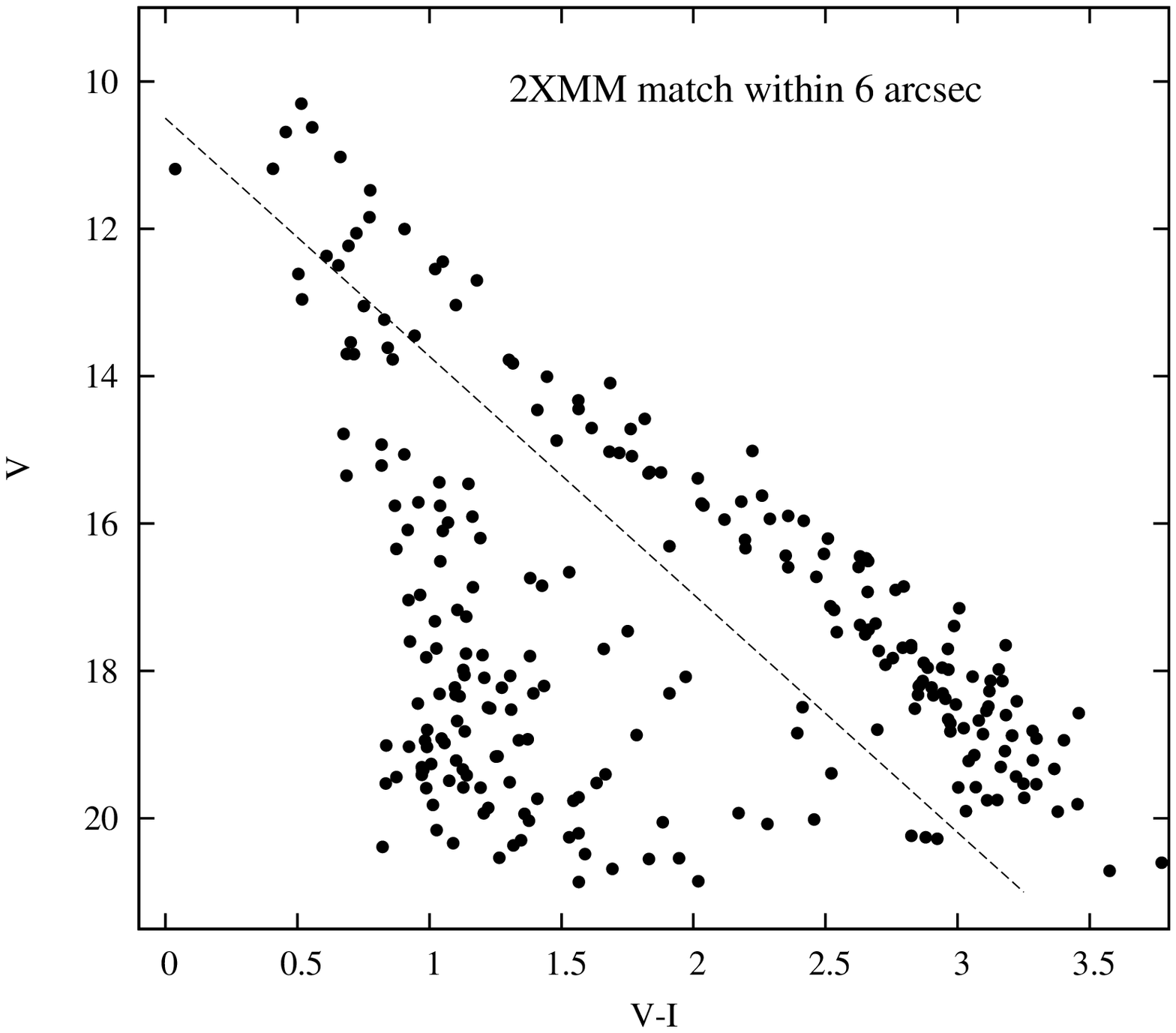}
\end{minipage}
\begin{minipage}[t]{0.45\textwidth}
\includegraphics[width=80mm]{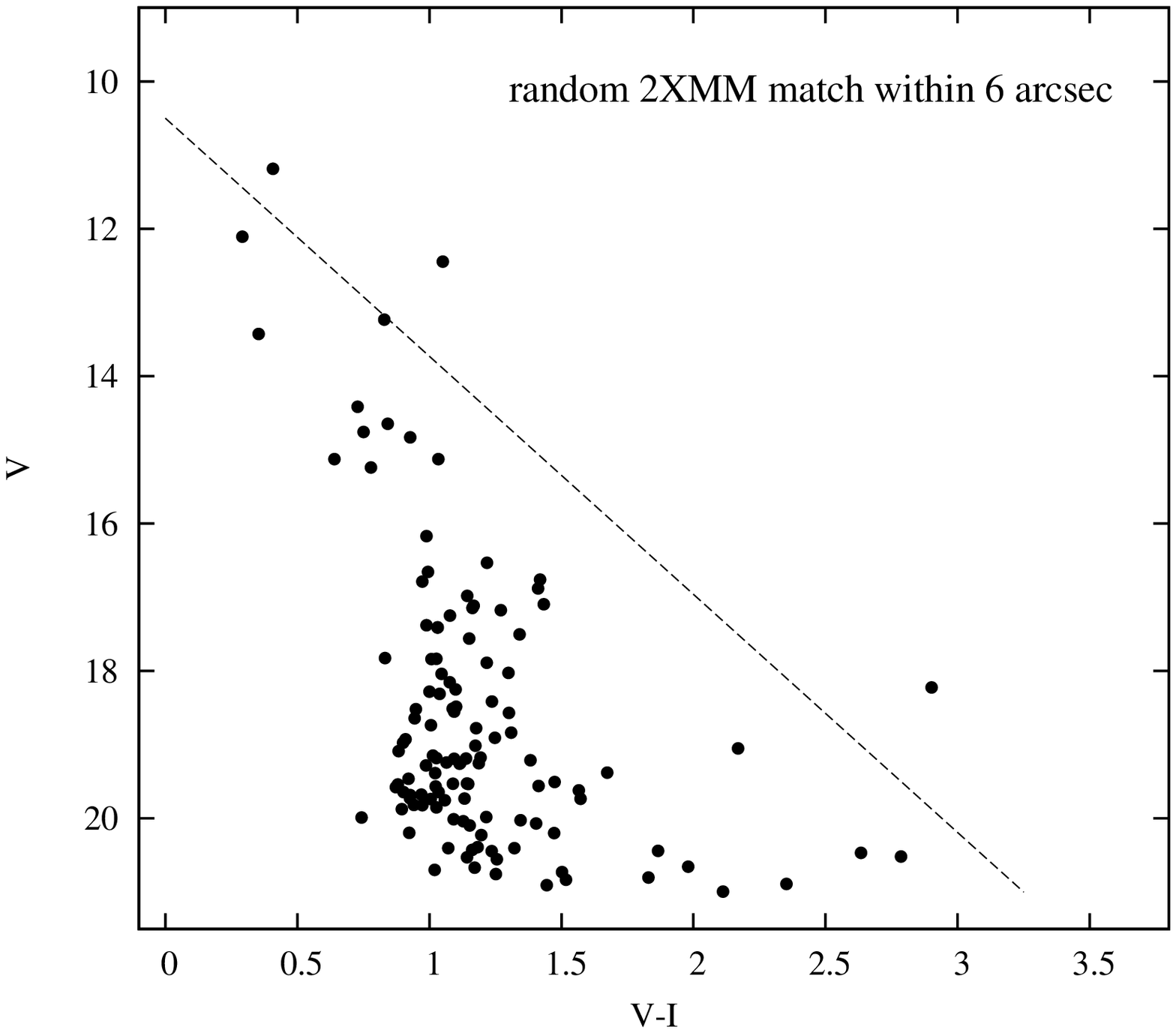}
\end{minipage}
\caption{(a) The CMD for unflagged optical sources
  correlated (up to a maximum of 6 arcsecs) with the merged 2XMM
  catalogue (see text).  (b) The likely distribution
  of spurious correlation in this CMD, obtained by offsetting all the
  X-ray sources by 30 arcsecs before performing the correlation. Only
  3 correlations above and to the right of the dashed PMS boundary
  defined in section~\ref{xmm} are expected to be spurious.}
\label{xraycmd}
\end{figure*}

The primary basis for assigning secure membership is 
the optical spectroscopy, which provides two good membership indicators
-- the RV and strength of the \lii~6708\AA\ feature.
The RVs of association members should lie close to a
uniform value, after allowing for some velocity dispersion, 
observational uncertainties and the possibility of close binaries. 
The abundance of lithium, and by extension
W[Li], is an age indicator for young, cool stars (see Jeffries 2006).
Lithium is depleted during the PMS phase in cool stellar envelopes
leading to a decline in W[Li] at a rate which depends on the mass and
hence effective temperature and colour of a star.
We refrain from using H$\alpha$ strength as a membership indicator
because we would like to comment later on the frequency with which strong,
accretion-related H$\alpha$ emission is seen.

Figure~\ref{lirv}a shows a plot of W[Li] versus RV.  Where two spectra
are available, the plotted information comes from the spectrum with the
highest signal-to-noise ratio. This plot shows a concentration of RV
values at 15--20\,kms\ and that objects with this RV tend to have the
largest values of W[Li].

Figure~\ref{lirv}b shows W[Li] as a function of $V-I$, again
plotting information from the best spectrum. The
symbols are coded to indicate which targets have an RV in the
range $11\leq $RV$ \leq 22$\kms. Also shown are loci defining the
observed upper and lower envelopes of W[Li] as a function of $V-I$ for
stars in the IC 2391, IC\,2602 and NGC 2547 open clusters (Randich
et al. 1997, 2001; Jeffries et al. 2003), which have ages in the range
30--50\,Myr. We also show a locus appropriate for a Li abundance
$A({\rm Li})=3.1$ (where $A($Li$)= 12 + \log N($Li)$/N($H$)$), which
corresponds approximately to an undepleted level of atmospheric Li. This
locus was calculated using atmospheric models and
colour-temperature relations described in Jeffries et al. (2003). 
These additional loci have been corrected by small amounts to
correspond to the reddening of the $\gamma$~Vel
association ($E(V-I)=0.055$, see section~\ref{fitting_ms}).

RV selection alone is unable to secure cluster
membership. Fig.~\ref{lirv}b shows that there are several stars with a
``membership RV'' but with a W[Li] inconsistent with being very young
stars ($<30$\,Myr). Membership or otherwise of the ``$\gamma$~Vel
association'' is assigned as follows:
\begin{enumerate}
\item
We consider all those stars in  Fig.~\ref{lirv}b
with W[Li] above the upper empirical locus for a 30--50\,Myr star
to be almost certain cluster members with one exception. The star 
ID 6-122 is at the boundary but distinct from many other objects with
$V-I \simeq 1.8$. Whilst this probably is a young star, its lack of a very
large W[Li] and anomalous RV make membership of the $\gamma$~Vel
association doubtful.
\item
There are five objects among the Li-rich members defined above where the
RV is not in the range 11--22\kms. One of these is a clear double-lined
spectroscopic binary (SB2, ID 4-34), one is a possible SB2 (ID 12-123)
and for the other three (ID 1-118, ID 2-256, ID 10-228) we were unable
to determine the RV because of the poor spectrum quality. The spectra
of these five stars are shown in Fig.~\ref{specplot}.  On the basis of their
W[Li] we consider all these to be association members.
\item
There are eight objects with $V-I<1.4$ that have a strong Li feature
which is at or below (by less than 1.5 error bars) the empirical upper
limit to W[Li] for a 30--50\,Myr old star. The usefulness of W[Li] as a
youth indicator declines for these warmer stars because the timescale
for Li-depletion is longer. However, because all eight have a RV
consistent with association membership we assume they are members.
\item The five bluest stars in the sample (with $V-I<0.5$) have only
  upper limits for W[Li] which might be consistent with undepleted
  Li. Three are also early-type stars for which we have been unable to
  ascertain an accurate RV. These three objects could be association members,
  but equally could be field objects and are classed as ``uncertain''. 
  The RV of the remaining two
  objects is incompatible with cluster membership, so we class them as non-members.
\item The rest of the objects which have only upper limits to their
  W[Li] are assumed to be non-members of the association. They either
  have a RV inconsistent with membership or lie far enough below the Li-rich members
  to make membership unlikely.
\end{enumerate}
A summary of our spectroscopic membership judgements is included as a column in
Table~\ref{specresults}. All together there are 32 stars considers as members, 49
as non-members and 3 with uncertain membership.

\subsection{Membership from the X-ray data}

\label{xraymembers}

\begin{table*}
\caption{(a) X-ray sources from the 2XMM catalogue which are correlated
  with 131 unflagged PMS optical counterparts in our photometric
  catalogue (those lying above the locus defined in
  Fig.~\ref{xraycmd}), appended with the details for 5 sources
  correlated with PMS counterparts that have flagged photometry, but which
  were included as spectroscopic targets. 
  Column 1 gives the IAU name of the 2XMM source,
  column 2 lists which observation the X-ray source parameters come
  from ($1=$ the observation beginning on MJD~52013, $2=$ the
  observation beginning on MJD~52033), column 3 gives the source
  detection likelihood, columns 4--9 gives the count rates and
  uncertainties in the 0.5-4.5\,keV range for the pn, m1 and m2
  instruments respectively (a zero here indicates the source was not
  found in that detector). Column 10 is the separation between X-ray
  source and optical counterpart,columns 11--12 are the optical
  counterpart identifiers from Table~2. Columns 13--14 are the RA and Dec
  (J2000) of the optical source, columns 15 and 16 give the $V$ and
  $V-I$ of the optical counterpart, column 17 is the average unabsorbed
  X-ray flux (0.5-4.5\,keV, erg\,cm$^{-2}$\,s$^{-1}$), column~18 the
  assumed $V$-band bolometric correction and column~19 the derived
  X-ray to bolometric flux ratio. (b) A similar table, but for those
  124 optical counterparts lying below the PMS selection locus shown in
  Fig.~\ref{xraycmd}.  These tables are only available in electronic
  form. A sample is shown below to illustrate the content.}
\begin{tabular}{c@{\hspace*{2mm}}c@{\hspace*{2mm}}c@{\hspace*{2mm}}c@{\hspace*{2mm}}c@{\hspace*{2mm}}c@{\hspace*{2mm}}c@{\hspace*{2mm}}c@{\hspace*{2mm}}c@{\hspace*{2mm}}c}
\hline
 (1)  & (2) & (3) & (4) & (5) &(6) &(7)&(8)&(9)&(10) \\
IAU Name & Obs & ML & pn & pn\_err & m1 & m1\_err & m2 & m2\_err &
Separation \\
         &     &
&(s$^{-1}$)&(s$^{-1}$)&(s$^{-1}$)&(s$^{-1}$)&(s$^{-1}$)&(s$^{-1}$)&
(arcsec) \\
\hline
2XMMp J080815.0-471537& 2&  4.7E$+$02& 0.00E$+$00& 0.0E$+$00& 2.00E$-$03& 2.6E$-$04& 2.50E$-$03& 2.2E$-$04& 0.9\\
2XMMp J080817.8-472246& 2&  1.2E$+$01& 1.56E$-$04& 1.8E$-$04& 3.59E$-$04& 9.4E$-$05& 0.00E$+$00& 0.0E$+$00& 1.6\\
2XMMp J080820.3-472026& 2&  2.9E$+$01& 9.86E$-$04& 4.8E$-$04& 2.51E$-$04& 7.9E$-$05& 2.69E$-$04& 8.1E$-$05& 2.2\\
2XMMp J080825.8-471639& 1&  1.7E$+$02& 2.64E$-$03& 3.7E$-$04& 9.02E$-$04& 1.8E$-$04& 1.06E$-$03& 2.1E$-$04& 1.5\\
\multicolumn{10}{c}{...}\\
\hline
\end{tabular}
\begin{tabular}{r@{\hspace*{2mm}}r@{\hspace*{2mm}}c@{\hspace*{2mm}}c@{\hspace*{2mm}}c@{\hspace*{2mm}}c@{\hspace*{2mm}}c@{\hspace*{2mm}}c@{\hspace*{2mm}}c@{\hspace*{2mm}}}
\hline
(11) & (12)& (13)& (14) & (15) & (16) & (17) & (18) & (19) \\
\multicolumn{2}{r}{Identifier} & RA & Dec & $V$ & $V-I$ & $f_{\rm x}$ &
BC$_{V}$ & $\log (L_{\rm x}/L_{\rm bol})$  \\ 
   &     &          &          &       &      &
(erg\,cm$^{-2}$\,s$^{-1}$) & (mag) &  \\
\hline
  6&  389& 122.06241& -47.26055& 15.947& 2.117& 4.10E$-$14&  -1.541& -3.06\\
  3& 1670& 122.07418& -47.37936& 18.599& 3.183& 3.79E$-$15&  -2.897& -3.57\\
  3&  576& 122.08422& -47.34054& 17.701& 2.963& 5.39E$-$15&  -2.584& -3.65\\
 12&  213& 122.10751& -47.27726& 16.206& 2.509& 1.79E$-$14&  -2.000& -3.50\\
\multicolumn{9}{c}{...}\\
\hline
\end{tabular}
\label{xraytable}
\end{table*}

Late-type PMS stars are strong X-ray emitters, with a ratio of X-ray to
bolometric luminosity, $L_{\rm x}/L_{\rm bol}$,
between $10^{-4}$ and $10^{-3}$. Field stars are not usually such
strong X-ray emitters unless they happen to be very young and rapidly
rotating or tidally locked in close binary systems. X-ray
emission is therefore an excellent way to identify PMS stars, although
because the bolometric luminosity of lower mass PMS stars is smaller, an
X-ray selected sample will become increasingly incomplete towards
fainter objects. The X-ray observations here are deep enough
that this limit is approached for much fainter objects than have optical
spectroscopy (see below), so an X-ray selected sample has the merit of clean
membership selection to lower masses.

The {\it XMM-Newton} field of view is roughly 30 arcminutes in diameter, so
only a subset of the photometric survey is covered by X-ray
observations. We looked for correlations between X-ray sources and
optical counterparts using a 6 arcsecond correlation radius. We
restricted the optical catalogue to those stars with
unflagged photometry and uncertainties in $V$ and $V-I$ less than
0.1 mag, finding 255 matches to 202 individual X-ray sources.  Randomly
offsetting the X-ray positions by 30 arcsecs reveals that about half of
these correlations are likely to be spurious because of the large
number of optical sources with faint magnitudes.  Fig.~\ref{xraycmd}
demonstrates that most spurious correlations will be with faint objects
having $V-I\simeq 1$. If we define a line on the CMD running from
(0.0,10.5) to (3.25,21), then the obvious PMS locus seen in the cool
part of the optical CMD and also the spectroscopic members (see
section~\ref{specmembers})
lie above this line, whilst the bulk of the
optical catalogue lies below it. Correlating the merged X-ray catalogue
against objects lying above the line accounts for 131 of the 
matches and only 3 of these are expected to be spurious 
(see Fig.~\ref{xraycmd} [right]).  The
conclusion is that the vast majority of X-ray selected PMS candidates
(i.e. those above the line) are genuine correlations and that spurious
correlations are predominantly fainter and bluer than the cluster members.
Tables~\ref{xraytable}a and~\ref{xraytable}b list the correlations between the 2XMM and
optical catalogues. Appended to Table~\ref{xraytable}a are details for
X-ray sources correlated with five flagged PMS counterparts that were
included as spectroscopic targets.

To assess whether the X-ray emission seen is characteristic of young
PMS stars we estimate the $L_{\rm x}/L_{\rm bol}$ ratio.
We have used the {\sc pimms} (version 3.9e)
tool from NASA's High Energy Astrophysics Science Archive Research
Center to convert EPIC count rates into unabsorbed X-ray fluxes in the
 0.5--4.5\,keV energy range. 
We assume a one temperature Raymond \& Smith (1977) thermal corona and
that the column density is $\simeq 3\times
10^{20}$\,cm$^{-2}$, which is appropriate for a reddening $E(B-V)\simeq 0.04$
(Bohlin, Savage \& Drake 1978). We assume an average coronal metallicty of $Z=0.4$,
since it is now well established that active PMS stars exhibit subsolar
coronal metallicities of this order (e.g. Scelsi et al. 2007). Finally, we used a
mean coronal temperature of $10^{7.0\pm 0.1}$\,K. This was
arrived at by considering the mean hardness ratio and standard deviation
(using the 0.5--1\,keV and 1--2\,keV bands) of 82 PMS objects
with good detections in the pn instrument. These assumptions
have little effect on the final estimated X-ray fluxes. The count-rate
conversion factor changes by: $^{+26}_{-10}\%$ if $\log T$ changes by
$\pm 0.3$; decreases by  1\% if $Z=1.0$; and by $^{+11}_{-6}\%$ if the
column density changes by $\pm 3\times 10^{20}$\,cm$^{-2}$.

\begin{figure}
\centering
\includegraphics[width=80mm]{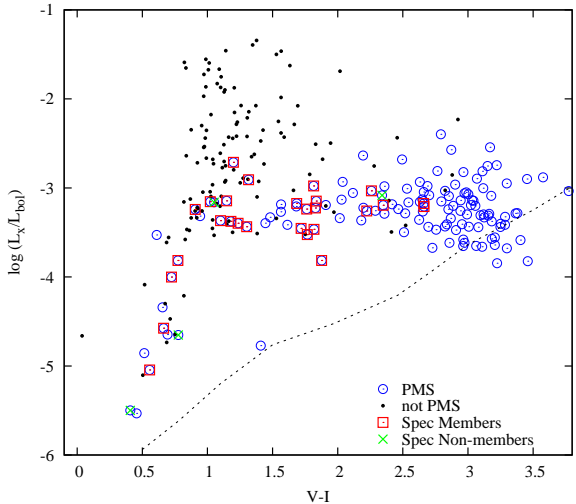}
\caption{X-ray to bolometric flux ratio, calculated assuming that all
  the X-ray sources have the coronal spectral parameters discussed in
  the text and have an extinction/reddening appropriate for the
  association. Stars which are PMS candidates based upon their
  photometry (i.e. above the dashed line in Fig.~\ref{xraycmd}) are
  indicated, as are those which have had their membership confirmed by
  spectroscopy or otherwise. The dashed line indicates the approximate sensitivity
  limit of the {\it XMM-Newton} observations for an object situated in the
  association PMS.}
\label{lxlbol}
\end{figure}

\begin{figure}
\centering
\includegraphics[width=80mm]{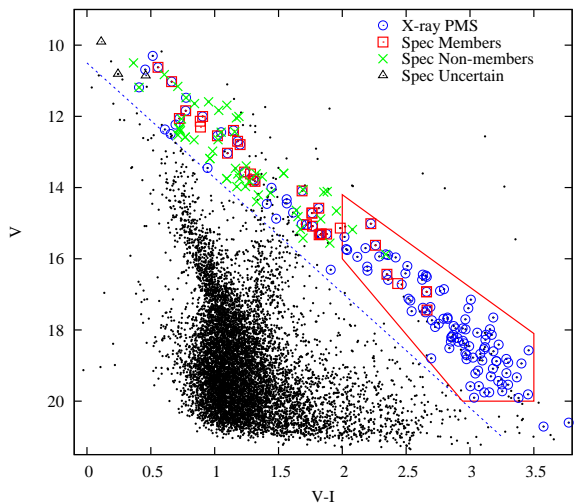}
\caption{The colour-magnitude diagram for objects from the photometric
  catalogue within the {\it XMM-Newton}
  field of view (small dots). 
  Additional symbols identify X-ray sources with a PMS optical
  counterpart, along with the membership status of objects we have 
  observed spectroscopically. The polygon defined by the solid line is
  used to select an almost complete and contamination-free sample of
  association members (see section~\ref{photomembers}).}
\label{bigcmd}
\end{figure}

The final conversion factors we use from pn (or MOS) count rates (in
the 0.5--4.5\,keV band) to unabsorbed X-ray fluxes (in the
0.5--4.5\,keV) band are $2.27\times 10^{-12}$\,erg\,cm$^{-2}$ (or
$6.06\times10^{-12}$\,erg\,cm$^{-2}$\,count$^{-1}$), where these conversion factors
apply to count rates evaluated over the whole point spread function of
the X-ray source (as supplied by the 2XMM catalogue).  Bolometric
fluxes were calculated using $V$ magnitudes, a bolometric correction
derived from the intrinsic $V-I$ (described in Naylor et al. 2002) and
assuming $E(V-I)=0.055$ and $A_{V}=0.131$ (see
section~\ref{agepms}). Table~\ref{xraytable} lists the average
fluxes, assumed bolometric corrections and $\log(L_{\rm x}/L_{\rm
bol})$ for both PMS and non-PMS X-ray sources.

Figure~\ref{lxlbol} shows the results and clearly demonstrates that the
X-ray selected PMS candiates, many of which have been confirmed by our
spectroscopy, show activity levels characteristic of PMS
populations -- a level of activity that increases with $V-I$ as
the objects develop deep convection zones, saturating at $L_{\rm
x}/L_{\rm bol}\simeq 10^{-3}$ for cool objects. Note there are
four spectroscopic non-members which are indistinguishable from
members on the basis of their X-ray activity. It is possible that these
are spurious correlations -- recall that $\sim 3$ were expected in the
PMS sample, and these are most likely to occur for $V-I<1.6$ where the
density of optical sources contaminating the PMS is greatest (see
Fig.~\ref{bigcmd}).  There is also one object (ID 12-75) with an X-ray
activity level much lower than other PMS candidates 
with $V-I\sim 1.5$ and is unlikely to be a cluster
member. We conclude that an X-ray selected sample of photometric PMS candidates
with $V-I>1.6$ is unlikely to be contaminated. Most of the remaining
(non-PMS) X-ray sources will either be active field dwarfs and
coronally active binary systems with $L_{\rm x}/L_{\rm bol}\leq
10^{-3}$ or, for the objects with larger X-ray to bolometric flux
ratios, active galaxies which are probably not even correlated with the
correct optical counterpart.

The minimum detectable unabsorbed X-ray flux in the {\it XMM-Newton}
observations is roughly $1.5\times10^{-15}$\,erg\,cm$^{2}$\,s$^{-1}$
when calculated using the conversion factors discussed above. The
dashed line in Fig.~\ref{lxlbol} shows how this translates into a
$L_{\rm x}/L_{\rm bol}$ detection threshold for a typical object that
is part of the X-ray selected PMS. As the X-ray luminosity functions of
PMS stars are not bimodal and there is a significant gap between the
$L_{\rm x}/L_{\rm bol}$ threshold and the detected PMS stars. 
There are 5 spectroscopically confirmed members which are apparently
not X-ray sources in Fig.~\ref{bigcmd}. In fact 3 of these are outside
the {\it XMM-Newton} field of view (ID 16-40, ID 6-216 and ID 8-24). 
One of them (ID 4-41) lies in the
wings of a bright X-ray source and is probably confused within it, and the
final object (ID 10-97) lies just over 6 arcsecs from an X-ray source
correlated with another spectroscopic member (ID 10-93). It is
likely that two separate X-ray sources have been merged by the {\it
XMM-Newton} spatial resolution and subsequent analysis.  We can
conclude that X-ray selection will be almost complete (where there is
X-ray coverage) for PMS stars with $V-I<2.8$, $V\simeq 18$, but
increasingly limited to the most active of stars that are even cooler.

\subsection{Photometric membership}
\label{photomembers}

Whilst an X-ray selected sample can be defined which is complete and
uncontaminated for $1.6<V-I<2.8$, this sample is spatially limited to
the 650\,arcmin$^2$ {\it XMM-Newton} field of view. The photometric
survey is more extensive so it makes sense to also define a
photometrically selected sample. Fig.~\ref{bigcmd} shows the
CMD within the X-ray field of view and
demonstrates that, for $2<V-I<3$, the vast majority of
photometric PMS candidates (those above the dashed line) are X-ray
sources with $L_{\rm x}/L_{\rm bol}>10^{-4}$ and therefore likely to be
very young stars. For hotter PMS candidates there are an increasing
proportion which are not X-ray sources or are
proven spectroscopic non-members. For cooler candidates a lack of
X-ray sensitivity probably leads to an increasing fraction of
non-detections. Note that Fig.~\ref{bigcmd} plots data for all of our
spectroscopic targets, some of which lie outside the {\it XMM-Newton} field of
view. In particular, there are eight spectroscopic members in
Fig.~\ref{bigcmd} that appear not to be X-ray sources. In fact, three
of these lie outside the {\it XMM-Newton} field (ID 6-216, ID 8-24 and
ID 16-40), while the remaining five have flagged photometry.

For the purposes of judging the spatial distribution of cluster members
we define a selection box as shown in Fig.~\ref{bigcmd} which should
result in a sample of PMS candidates with $V-I>2$ that are nearly
uncontaminated and complete for $V\leq 20$, the approximate
completeness limit of the photometric survey. To put this on a
quantitative basis, there are 24/116 objects in this photometric
selection box that are {\em not} X-ray sources or are spectroscopic
non-members. However, the majority of these have $V>18$ and have
probably just not been detected in X-rays {\it yet}. This means that
the contamination level is $<0.037$\,arcmin$^{-2}$ if we assume all the
X-ray sources in the selection box are members of the association.
Applying the selection box to the full photometric catalogue defines
a sample of 319 unflagged photometric members with $V<20$.

\subsection{Membership of bright stars}

\label{memberbright}

\begin{figure}
\includegraphics[width=80mm]{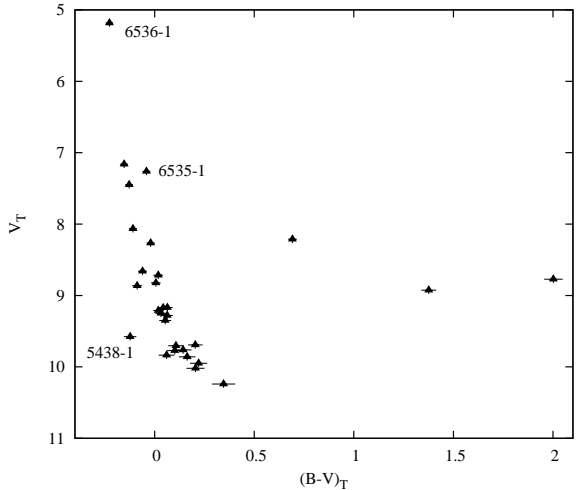}
\caption{
The colour-magnitude diagram for objects in the Tycho Catalogue within 45
arcminutes of $\gamma$ Vel and which satisfy a proper motion criterium
(see text).
}
\label{pm_sel}
\end{figure}

The saturation level of our new CCD photometry is $V\simeq 10.5$. To
get information on brighter stars we consult the Tycho-2 catalogue (H{\o}g
et al. 2000) which contains $V$ and $B-V$ magnitudes in the Tycho
system along with precise positions and proper motions.

To define a sample of bright association candidates
we take those stars within 45 arcminutes of $\gamma^2$ Vel 
that satisfy two further criteria.  (i) Their proper motions are
within three times their error bar of the mean proper motion of the
Vela OB2 association derived from Hipparcos
data by de Zeeuw et al. (1999). We add 1 mas/year in quadrature to the proper motion
uncertainty (corresponding to $\simeq 1.5$\,km\,s$^{-1}$ in tangential
motion) to account for any small velocity dispersion (see
section~\ref{kinspace}).  
(ii) They have photometric uncertainties of $<0.05$ mag in $V$ and $B-V$
(in the Tycho system).  This proper motion and position
selection results in the catalogue of 29 stars shown in
Fig.~\ref{pm_sel}.  There are three objects with $B-V>0.5$
(Tyc-8140-4503-1, Tyc-8140-2697-1, Tyc-8140-3771-1) which clearly stand out as
anomalous. These are most likely unrelated
background giants -- they would have to be $>300$\,Myr old if they were
red giants at the distance of the association.

The majority of the stars selected form a sequence in colour-magnitude
space (Fig.~\ref{pm_sel}), and share the same proper motion.  It is
quite likely they are a distinct, co-eval population. Since they
share the same position on the sky as our PMS sample, it
is also likely that both groups are part of the same population.
Indeed, it is almost inconceivable that the two are not the same
population, since were they not, the question would be where in the
CMD are the higher-mass stars which correspond to
the large number of low-mass stars we have found (see section~\ref{massfunction})?

\section{Properties of the association around Gamma Vel}

\subsection{Kinematic and spatial coherence}

\label{kinspace}

\begin{figure}
\includegraphics[width=80mm]{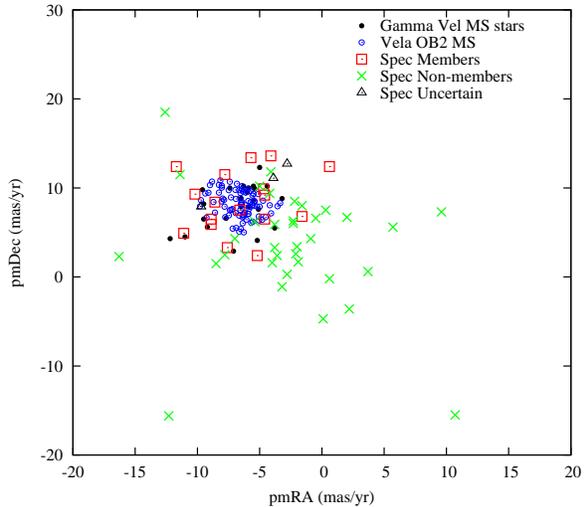}
\caption{Proper motions for the various stellar samples discussed in
  the text. The $\gamma$~Vel and Vela OB2 main-sequence proper motions come
  from the Tycho and Hipparcos catalogues respectively. The proper
  motions for the spectroscopic targets are good quality measurements 
  from the NOMAD catalogue.
}
\label{pmplot}
\end{figure}

Figure~\ref{lirv} shows that the Li-rich stars are closely
grouped in RV. If we take the 27 spectroscopically confirmed PMS candidates that
have $11\leq RV \leq 22$\,km\,s$^{-1}$,
then the mean association heliocentric RV is 17.1\,km\,s$^{-1}$ with a
standard deviation of 2.5\,km\,s$^{-1}$. This standard deviation must
be a strong upper limit to the 1-dimensional velocity dispersion in
the cluster. We estimate that the RVs are uncertain at the level of
2--3\,km\,s$^{-1}$, which must account for the majority of the observed
dispersion. No external RV standards (other than the Sun)
were observed during our spectroscopy run. Experience from other
datasets suggests that the uncertainty in the mean RV will be dominated
by systematic effects at the level of $\sim \pm 1$\,km\,s$^{-1}$, 
rather than the smaller statistical uncertainty in the mean.

The systemic RVs of $\gamma^{1}$ and $\gamma^{2}$~Vel are $9.7\pm 1.0$
and $7\pm 23$\,km\,s$^{-1}$ respectively (Hernandez \& Sahade 1980;
Schmutz et al. 1997). The very small error bar on the RV of
$\gamma^{1}$~Vel should be treated with caution. These were
photographic spectra with RVs averaged over many spectral lines. It is
unclear how the line centres were determined or whether there was
significant variation between lines. No systematic study of the RVs of
other early-type stars near $\gamma$~Vel has been reported.

Proper motions for the Vela OB2 main sequence (MS) stars and the bright
Tycho-selected MS stars around $\gamma$~Vel are available from the
Hipparcos and Tycho catalogues respectively, and have typical
uncertainties of 0.8 and 1.5 milli-arcsec/year in each coordinate. We
searched the {\sc nomad} catalogue (Zacharias 2005) for proper motions
for fainter sources in our optical catalogue. Good quality proper
motions (flagged as of astrometric standard) are available for 58 of
the 84 stars with spectroscopic membership determination.  All of these
have $V<15.5$ and uncertainties of $\sim 5$ milli-arcsec/year in each
coordinate.

\begin{figure*}
\centering
\begin{minipage}[t]{0.45\textwidth}
\includegraphics[width=80mm]{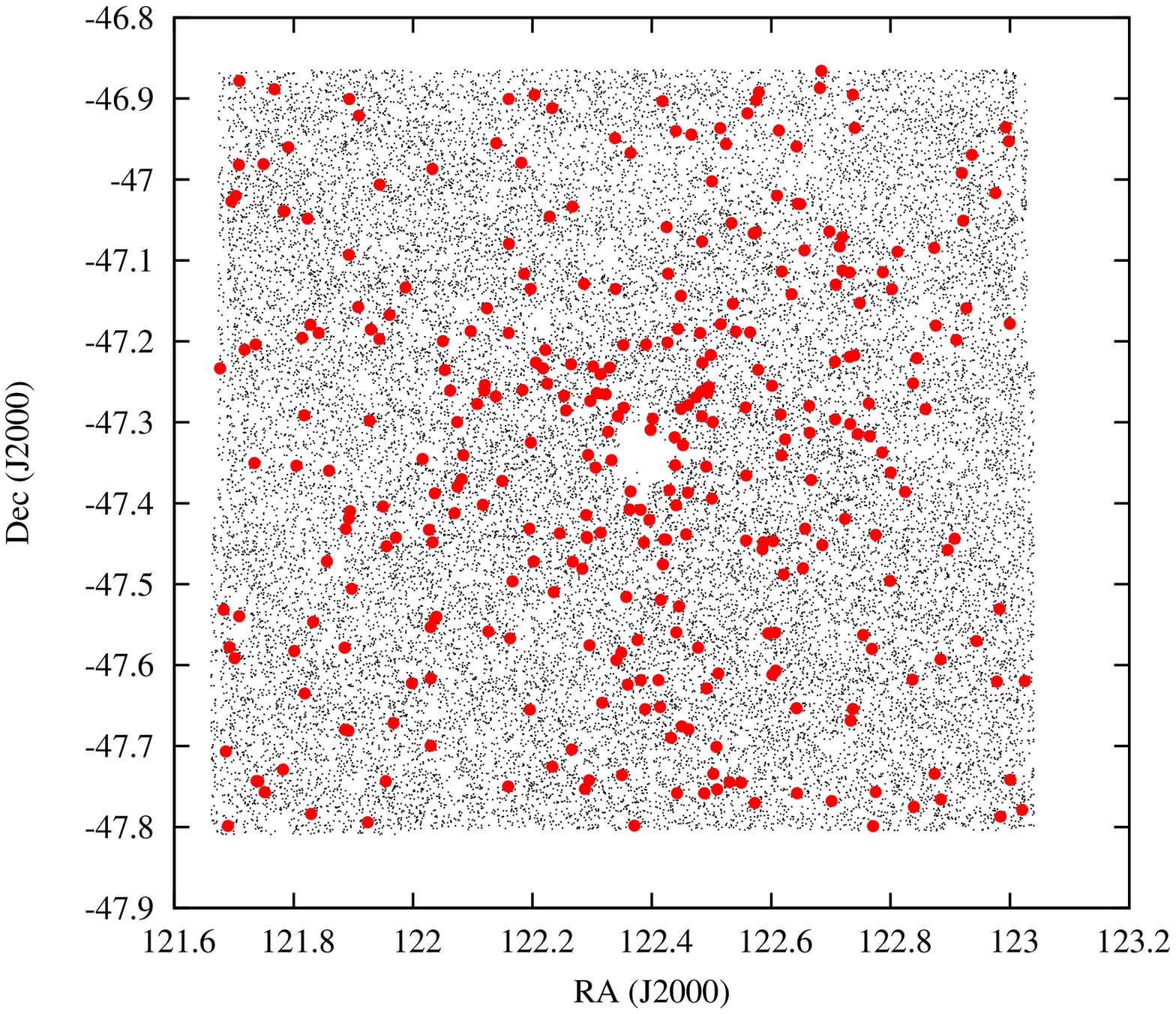}
\end{minipage}
\begin{minipage}[t]{0.45\textwidth}
\includegraphics[width=80mm]{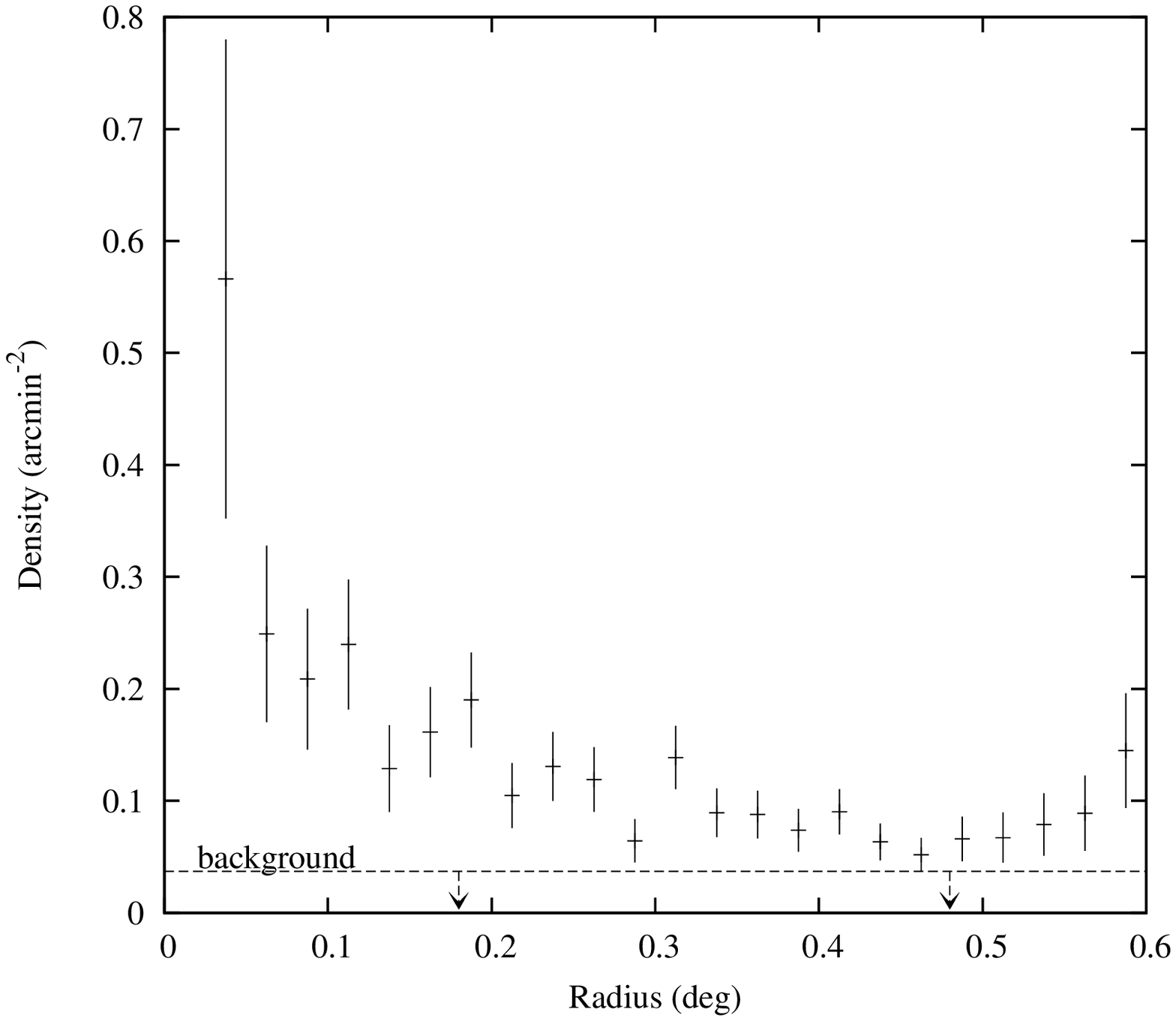}
\end{minipage}
\caption{(a) Large spots show the locations of the low-mass PMS
  photometric candidates defined in section~\ref{photomembers}. The
  small dots show the quasi-uniform distribution of the much larger
  number of stars with similar brightness (but bluer colours). The ``hole''
  in the middle is caused by the lack of good photometry for stars that
  are close to $\gamma^2$~Vel. (b) The radial distribution of the
  photometric PMS candidates, normalised by the radial distribution
  (centred on $\gamma^2$~Vel) of the comparison sample with similar brightness.
  The central bin is missing because no good photometry was obtained
  close to $\gamma^2$~Vel. The horizontal dashed line shows an upper
  limit to the level of contamination in the sample.
}
\label{spatial}
\end{figure*}

\begin{table*}
\begin{tabular}{lcccccc}
\hline
Group   & Nstars & Avg. uncertainty & \multicolumn{2}{c}{pm RA (mas/yr)} &
\multicolumn{2}{c}{pm Dec (mas/yr)} \\
        & &(mas/yr)       &  mean & std. dev. & mean & std. dev. \\
\hline
$\gamma^2$ Vel & 1 & -- & $-5.9$ & 0.5 & $+9.9$ & 0.4 \\
Vela OB2 MS & 85 & 0.8 & $-6.6$ & 1.3 & $+8.1$ & 1.4 \\
$\gamma$ Vel MS& 24 & 1.5 & $-6.8$& 2.2 &$+7.9$ & 2.4 \\
Spectroscopic Members & 17 & 4.4 & $-6.5$ & 3.2 & $+8.5$ & 3.3 \\
Spectroscopic Non-Members & 38 & 4.1 &$-3.2$ & 6.8 & $+4.1$& 11.5 \\
\hline
\end{tabular}
\caption{A summary of the the proper-motion properties of various star
  samples (see text).}
\label{pmtable}
\end{table*}

The proper motions are plotted in Fig.~\ref{pmplot} and the mean and
standard deviations of the proper motion for each of these stellar
groups are reported in Table~\ref{pmtable}. Note that the MS stars
around $\gamma$~Vel and the Vela OB2 stars have been selected on the
basis of their similar proper motions, yet may still contain some
non-members.  The key point is that the spectroscopic members are
indistinguishable from $\gamma^2$~Vel, the Vela OB2 MS stars and the MS
stars around $\gamma$~Vel on the basis of their proper-motions. The
mean proper motions agree to within their small statistical
uncertainties and the observed standard deviations are close to the
statistical uncertainties, suggesting that any internal scatter in the
proper-motions is small.  Quantitatively, the internal scatter is
probably less than 1.5 milli-arcsec/year for the MS stars around
$\gamma$~Vel ($<2.5$\,km\,s$^{-1}$ at the distance of $\gamma$~Vel, see
section~\ref{fitting_ms}) and consistent with zero for the
spectroscopic members. On the other hand the spectroscopic non-members
show a larger dispersion and a different mean proper-motion. On that
basis it seems likely, although not conclusive, that the 3 stars
with ``uncertain'' status (triangles in Fig.~\ref{pmplot}) are members.
Given sufficiently precise proper motions (of order 2 milliarcsec/year)
it should be possible to separate cluster and field populations with
some confidence. However the proper motions we have are limited to
relatively bright stars ($V<16$) and are quite incomplete, so cannot be
effectively used for that purpose in this paper.

To examine the spatial structure of the low-mass association we use the
photometrically defined sample of PMS candidates
(section~\ref{photomembers}). This covers the entire photometric survey
and should have reasonably uniform spatial sensitivity apart from
the region immediately around $\gamma^2$ Vel itself (see
section~\ref{photometry}). 

Figure~\ref{spatial}a shows the spatial distribution of the 319 photometric
candidates from the selection box in Fig.~\ref{bigcmd}. The radial distribution of
these objects is shown in Fig.~\ref{spatial}b, centred on
$\gamma^2$~Vel. We have made a
correction (which is only significant for the first two bins) for
spatial incompleteness in the survey, by normalising to the radial
distribution of stars selected from a comparison box in the $V,V-I$ CMD
with $V-I<2$ and $16<V<20$. An upper limit to the level of
contamination (derived from $<24$ non-members in the 650 arcmin$^{2}$
{\it XMM-Newton} field of view) is also shown.

A clear central concentration in the spatial distribution of
photometric association candidates is evident. But it is also apparent
that the edge of the spatial distribution has not been reached, because
the spatial density of candidates members is still significantly higher
than the upper limit to the background level even at radii of 0.5--0.6
degrees.

\subsection{Distance and Extinction for a main sequence sample}
\label{fitting_ms}

\begin{figure}
  \begin{center}
      \includegraphics[width=80mm, clip=true]{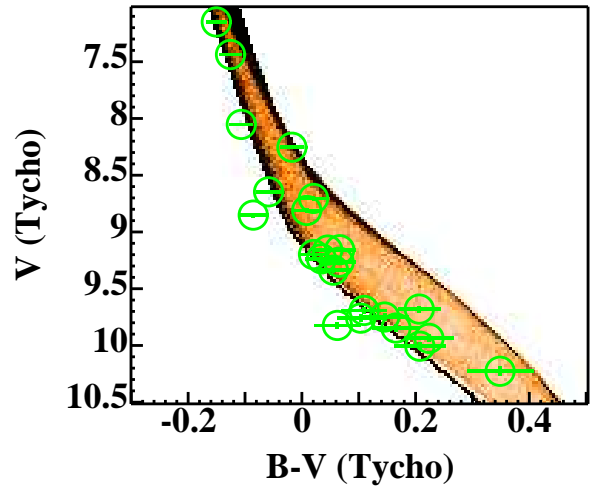}
  \end{center}
  \caption{ 
    The best fitting model (colour scale) to the dataset of Figure \ref{pm_sel}.
  }
  \label{gamvel_fit}
\end{figure} 

\begin{figure}
  \begin{center}
      \includegraphics[width=80mm,clip=true]{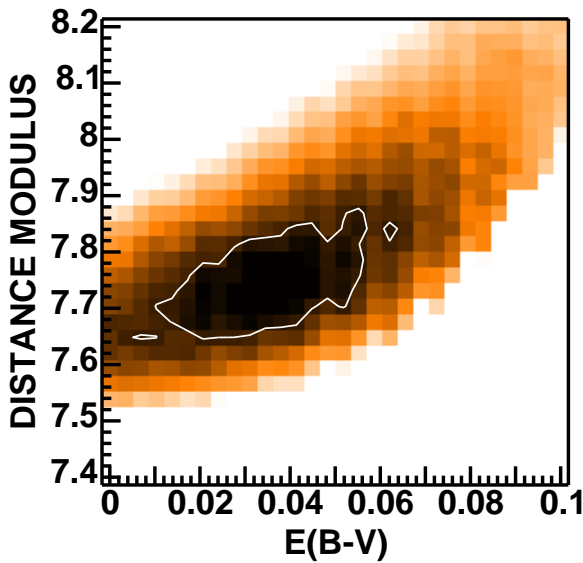}
  \end{center}
  \caption{ 
    $\tau^2$ as a function of distance modulus and extinction for the
    data and model of Figure \ref{gamvel_fit}.  The contour shows the
    68 percent confidence region.
  }
  \label{gamvel_grid}
\end{figure} 

The low-mass stars identified as young association members based on
their RVs, lithium abundance and X-ray emission, are PMS stars which
fade as they contract.  The sequence they form moves to fainter
magnitudes with age, resulting in a strong age-distance degeneracy if
we were to try and fit that sequence with model isochrones (e.g. Naylor
\& Jeffries 2006).  Conversely, high-mass stars around $\gamma$ Vel
have already reached the MS, with almost age-independent positions in a
CMD (for a quantitative discussion see Mayne \& Naylor 2008).  By
measuring the distance to these we can pin down the distance to the
low-mass objects and hence get a better age estimate.

Later we shall show that the MS sample around $\gamma$~Vel is at a
similar distance to the majority of the Vela OB2 association, and that
$\gamma$ Vel itself also lies at this distance, but for the moment one
should be quite clear that fitting the proper-motion-selected MS sample
from section~\ref{memberbright} does not actually measure the distance
to either $\gamma$~Vel or Vela OB2.

We fitted the MS dataset from section~\ref{memberbright} 
using the Geneva-Bessell models described in
Mayne \& Naylor (2008).  These are based on the Geneva
isochrones (Lejeune \& Schaerer 2001) but use bolometric corrections
and $T_{\rm eff}$-colour conversions from Bessell et al. (1998), assuming
the magnitude of Vega is zero in both $V$ and $B$.  These are
converted from the conventional $BV$ system to the Tycho system using
formulae in Bessell (2000).  Using these models to simulate roughly a
million stars allowed us to create a probability density
for the position of a star in the CMD plane (the colour scale in Figure
\ref{gamvel_fit}).  We then used the $\tau^2$ statistic (Naylor \&
Jeffries 2006; Naylor in prep.) 
to compare the model and data for any value of
distance modulus and extinction, and find best-fitting values.

Before commencing we removed the three objects with
$B-V>0.5$ (see section~\ref{memberbright}). In addition we removed the star
(Tyc-8140-5438-1) which lies too blue of the MS (see Fig.~\ref{pm_sel}).  Finally, our
bolometric correction runs out at $V_0 \approx -1.7$, which for
reasonable distances places star Tyc-8140-6536-1 quite close to the brightest
available model. We therefore removed this point as well.  We fitted
the resulting dataset to a 7\,Myr isochrone using the colour-dependent
extinction vectors derived for the Tycho system in Mayne \& Naylor
(2008) and assuming a binary fraction of 50 per cent.  Mayne \& Naylor
(2008) show that the best-fit extinction and distance are not very
dependent on the assumed isochronal age, because the fitted stars are much
fainter than the MS turn-off. If we fit the resulting
dataset without any clipping of high $\tau^2$ datapoints (see Naylor \&
Jeffries 2006) we obtain a $P_r(\tau^2)$ of 0.15 with a distribution of
$\tau^2$ which suggests that star Tyc-8140-6535-1, a bright star just to the
red of the sequence should also be removed from the fit.  Removing this
point results in a $P_r(\tau^2)$ of 0.64, and the fit shown in Figure
\ref{gamvel_fit}.  We considered the possibility that both Tyc-8140-6535-1
and Tyc-8140-5438-1 should be included because they represent a real spread
in distance. However, these stars lie over a magnitude (in opposite
directions) from the sequence, making such an interpretation unlikely.

The plot of $\tau^2$ as a function of distance and extinction
(analogous to a $\chi^2$ space), is shown in Figure \ref{gamvel_grid}.
Since  we have a reasonable $P_r(\tau^2)$ value we calculate
parameter uncertainties as described in Naylor (2008 in
prep), without any need to increase the uncertainties for the
individual  datapoints. Doing so results in the confidence
interval~shown in Fig.~\ref{gamvel_grid}.  The lowest $\tau^2$
(corresponding to the most likely parameter values) occurs at an
absolute distance modulus of $7.76\pm 0.07$ mag and $E(B-V)$ of
$0.038\pm0.016$ mag, where the uncertainties are 68 per cent confidence
levels for one parameter of interest.

\subsection{The Distance to Vela OB2}

\label{velaob2}

\begin{figure}
  \begin{center}
    \begin{tabular}{c}
      \includegraphics[width=80mm, clip=true]{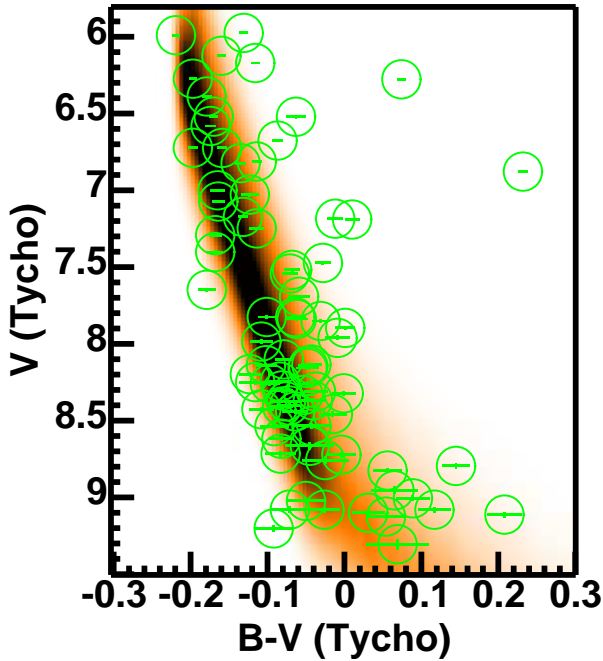}
    \end{tabular}
  \end{center}
  \caption{ 
    The best fitting model (colour scale) to the 77 Vela OB2
    association members defined by de Zeeuw et al. (1999) and which
    have $V_{T}>6$ and $(B-V)_T <0.5$.  The model includes a distance
    modulus spread of 0.3 mag.
  }
  \label{dezeeuw_fit}
\end{figure} 

\begin{figure}
\includegraphics[angle=90, width=80mm, clip=true]{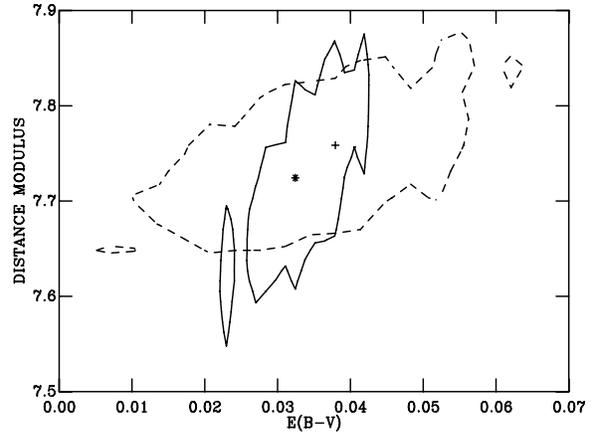}
\caption{
The 68 percent confidence limits in intrinsic distance modulus and
extinction
for the de Zeeuw sample (solid contour) and
the MS sample around $\gamma$ Vel (dashed contour). The asterisk and
cross respectively mark the best-fit values for these samples.
}
\label{conf_regions}
\end{figure}

To establish a distance to the surrounding Vela OB2 association on the
same basis we fitted 
Tycho photometry of the members given by de Zeeuw et al. (1999). These
stars were chosen by de Zeeuw et al. on the basis of photometry
and proper motions and lie in a region more than 10 degrees in
diameter. We again limited ourselves to stars
with $V_T>6$, due to the limitation of our calibration, and
ignored objects with $(B-V)_T>0.5$, since these are either non-members,
or perhaps evolved stars.

Such a fit, using a clipping threshold of $\tau^2$=10, clips 19 of the 77 available
datapoints, and still only achieves $P_r(\tau^2)=0.00004$.  The
absence of a clearly defined sequence implies that there may be a
spread in distance.  We modelled this as a spread in distance modulus,
by convolving a Gaussian with the expected distribution in
colour-magnitude space.  This procedure cannot be carried out in a
statistically rigorous fashion, since adding the distance modulus
spread changes the number of datapoints which are clipped out.  This
ruins any chance of a strict statistical comparison between models with
different distance spreads, but in any case clipping is not a well
defined process in statistical terms.  However, if we introduce a
Gaussian spread in distance modulus with $\sigma=0.3$ mags, then 
$P_r(\tau^2)$ rises to about 0.52 for only 13 clipped
datapoints.  Furthermore the distribution of $\tau^2$ for the unclipped
points is similar to that predicted from the model.
Introducing the distance spread has little effect on the
derived best-fit parameters, changing the distance modulus and
$E(B-V)$ from 7.65 and 0.028 mag respectively to 7.68 and 0.036 mag.  

To derive a reasonable estimate of the uncertainties for the model with
a distance spread we repeat the $\tau^2$ grid search with the clipped
datapoints removed. We find a best-fitting $E(B-V)=0.032\pm0.009$\,mag and
absolute distance modulus of $7.72\pm0.08$ mag.  
The
best-fitting model (with a distance modulus spread of 0.3 mags) is
shown with the data overlayed in Figure \ref{dezeeuw_fit}.
The 68 percent confidence
contour is plotted on Fig.~\ref{conf_regions}. It is clear that the
sample of MS stars around $\gamma$ Vel and the de Zeeuw Vela OB2 sample
have an indistinguishable distance modulus and extinction at this level.

The distance modulus we have found for the Vela OB2 sample is closer
than the $(8.06 \pm 0.07)$\,mag found by de Zeeuw et al. (1999). We
have no definitive explanation for this.  There have been problems with
reported mean Hipparcos parallaxes to some clusters and associations
(notably the Pleiades -- see Pinsonneault et al. 1998) probably due to
correlated parallax errors over small regions on the sky. Another
factor may be the accuracy of the correction de Zeeuw et al. have
applied to account for biases introduced by discarding objects with
negative parallax and incompleteness for stars with fainter magnitudes.
A further possibility is a departure from our assumption of solar
metallicity. This is discussed further in section~\ref{discussion}.

The suggestion of a $\pm 0.3$~mag spread in distance modulus is
not altogether surprising. The stars defined as the Vela OB2 association by de
Zeeuw et al. (1999) occupy a roughly circular area of about 12 degrees
diameter. If the association is spherical then the front-to-back
distance spread would be of order $\pm 0.2$ mag. There also remains the
possibility of small star-to-star reddening variations which if modelled as a 
vertical displacement in the CMD could mimic a significant distance spread.

\subsection{The Age of the Pre-Main-Sequence Sample}

\label{agepms}

\subsubsection{Low-mass PMS model isochrones}

To determine an age for the PMS sample we adopt the best fitting
distance modulus and extinction from fitting the MS stars whose
projected positions lie around $\gamma$ Vel (see section
\ref{fitting_ms}).  Extinction and reddening in terms of $A_V$ and $E(V-I)$
are calculated for a star with a typical intrinsic PMS colour as
$A_V=3.48 E(B-V)=0.131$ and $E(V-I)=1.44 E(B-V)=0.055$ (see Bessell et
al. 1998).  

\begin{figure*}
\centering
\begin{minipage}[t]{0.45\textwidth}
\includegraphics[width=80mm]{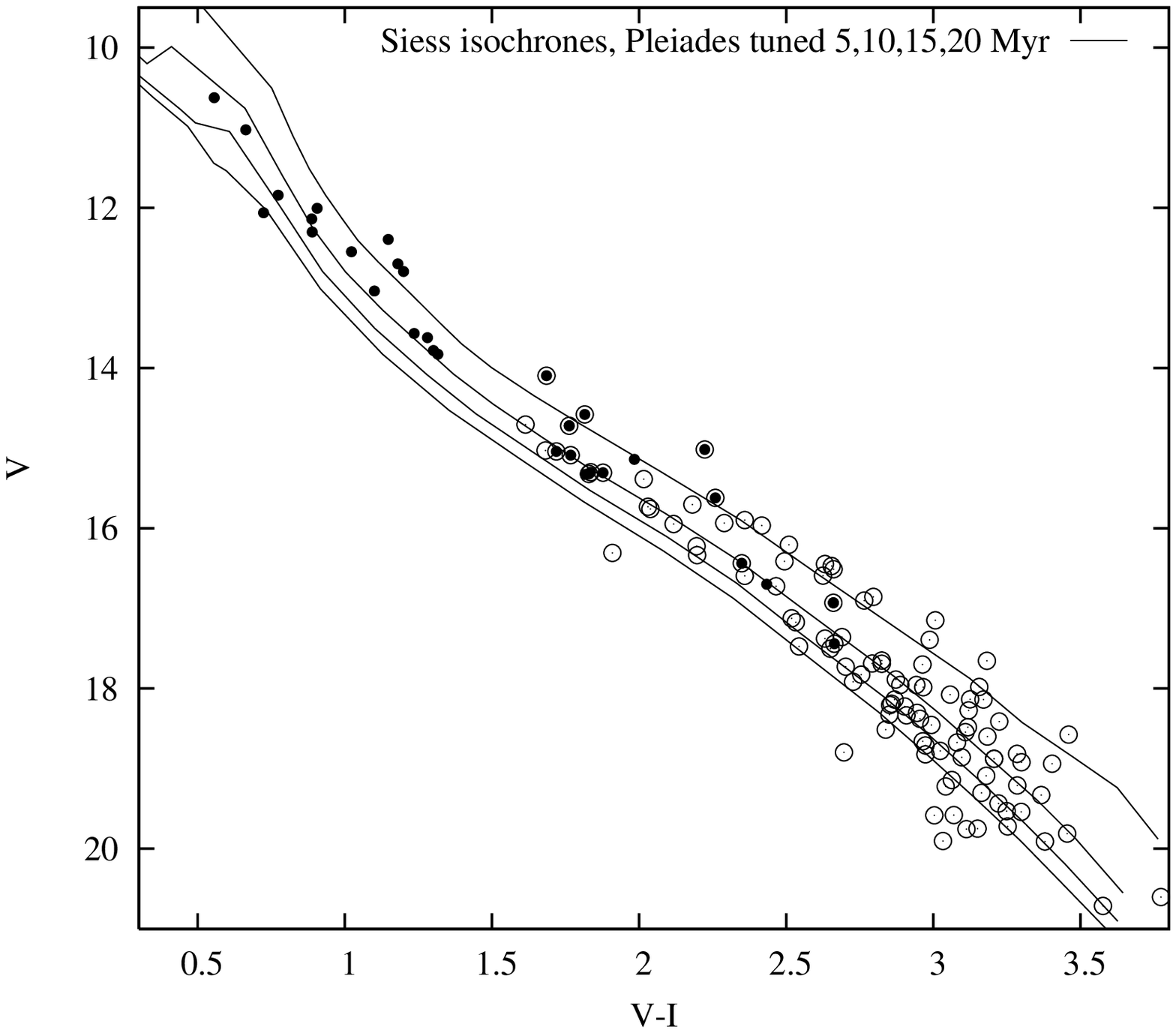}
\includegraphics[width=80mm]{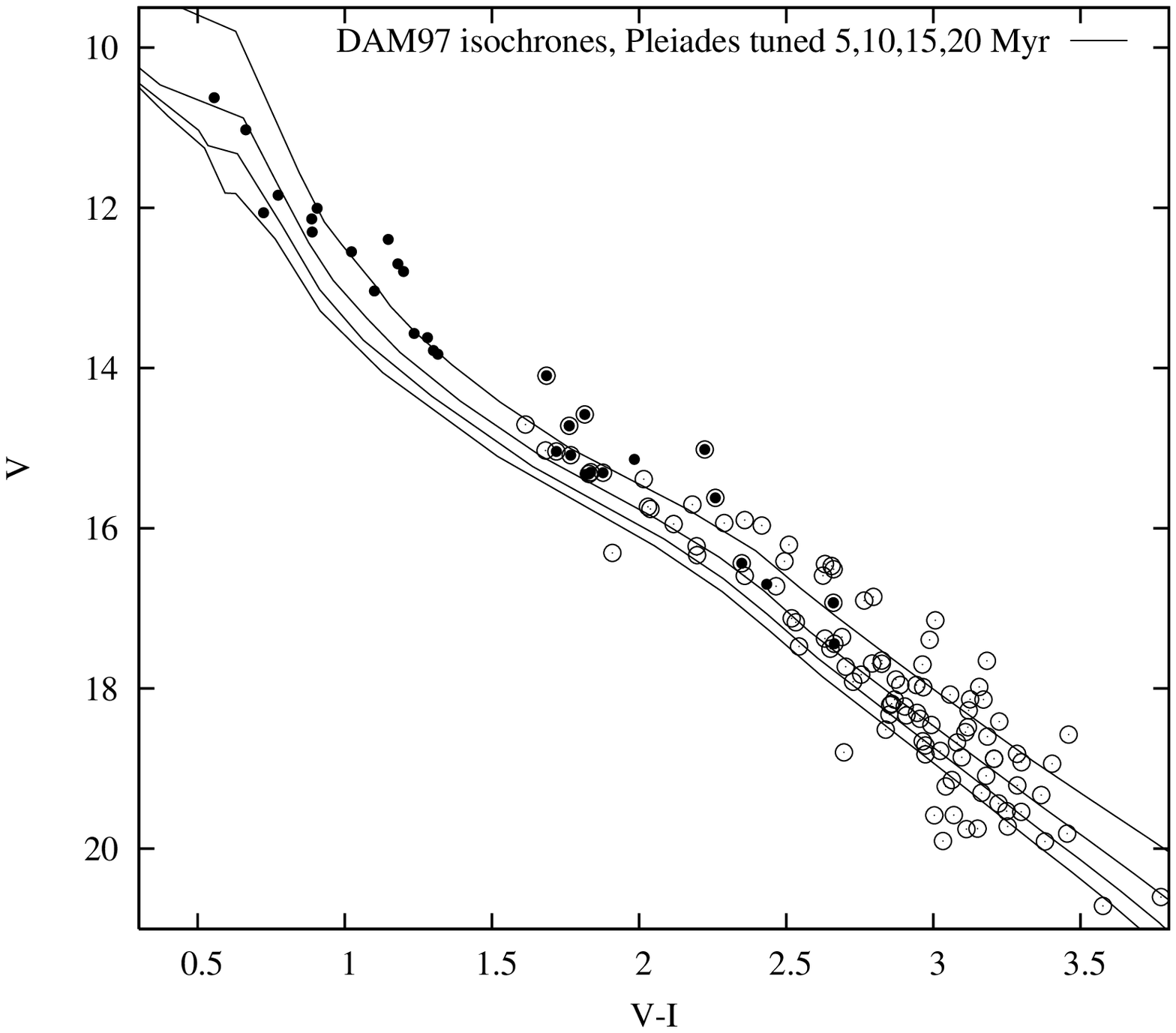}
\end{minipage}
\begin{minipage}[t]{0.45\textwidth}
\includegraphics[width=80mm]{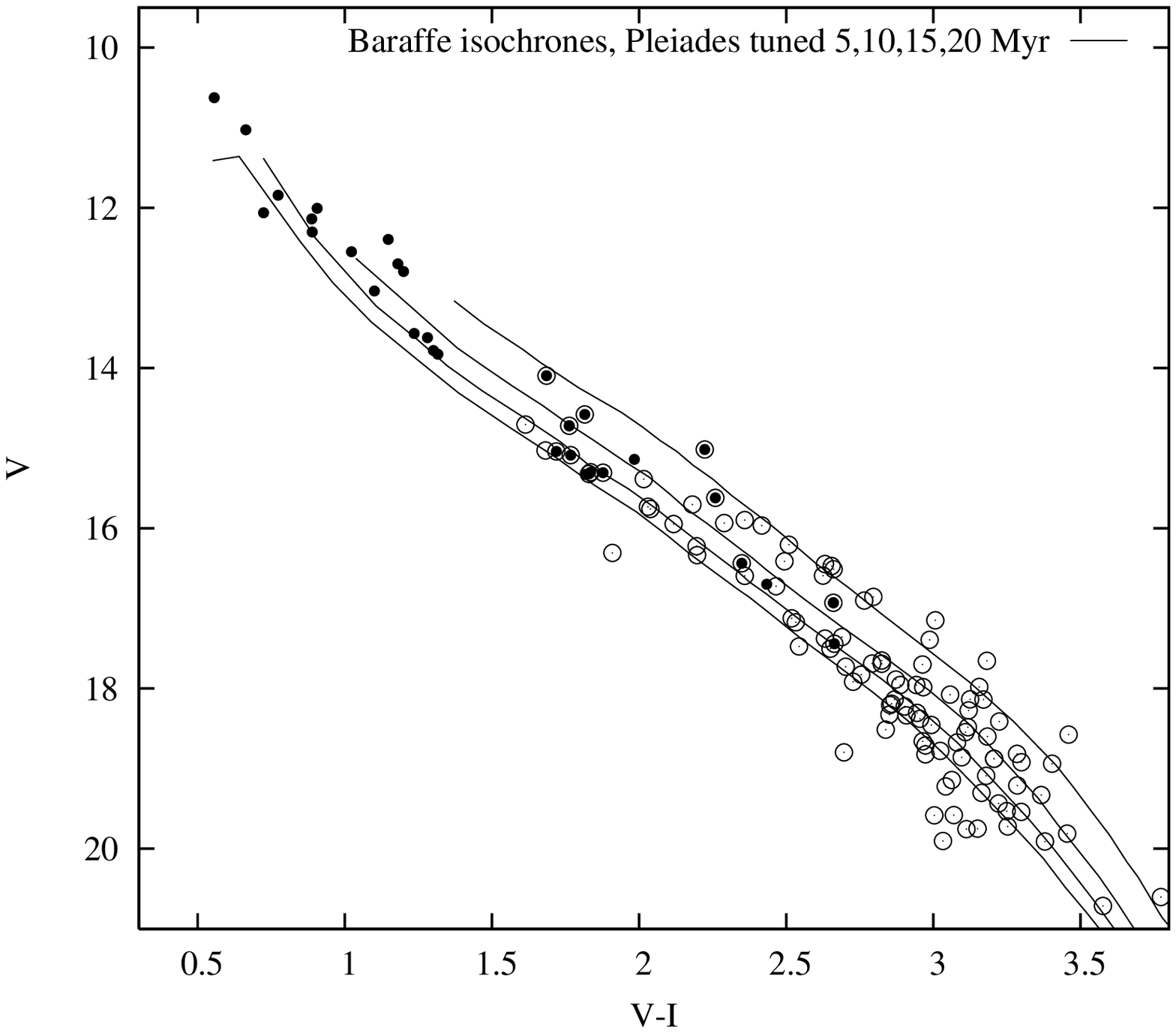}
\includegraphics[width=80mm]{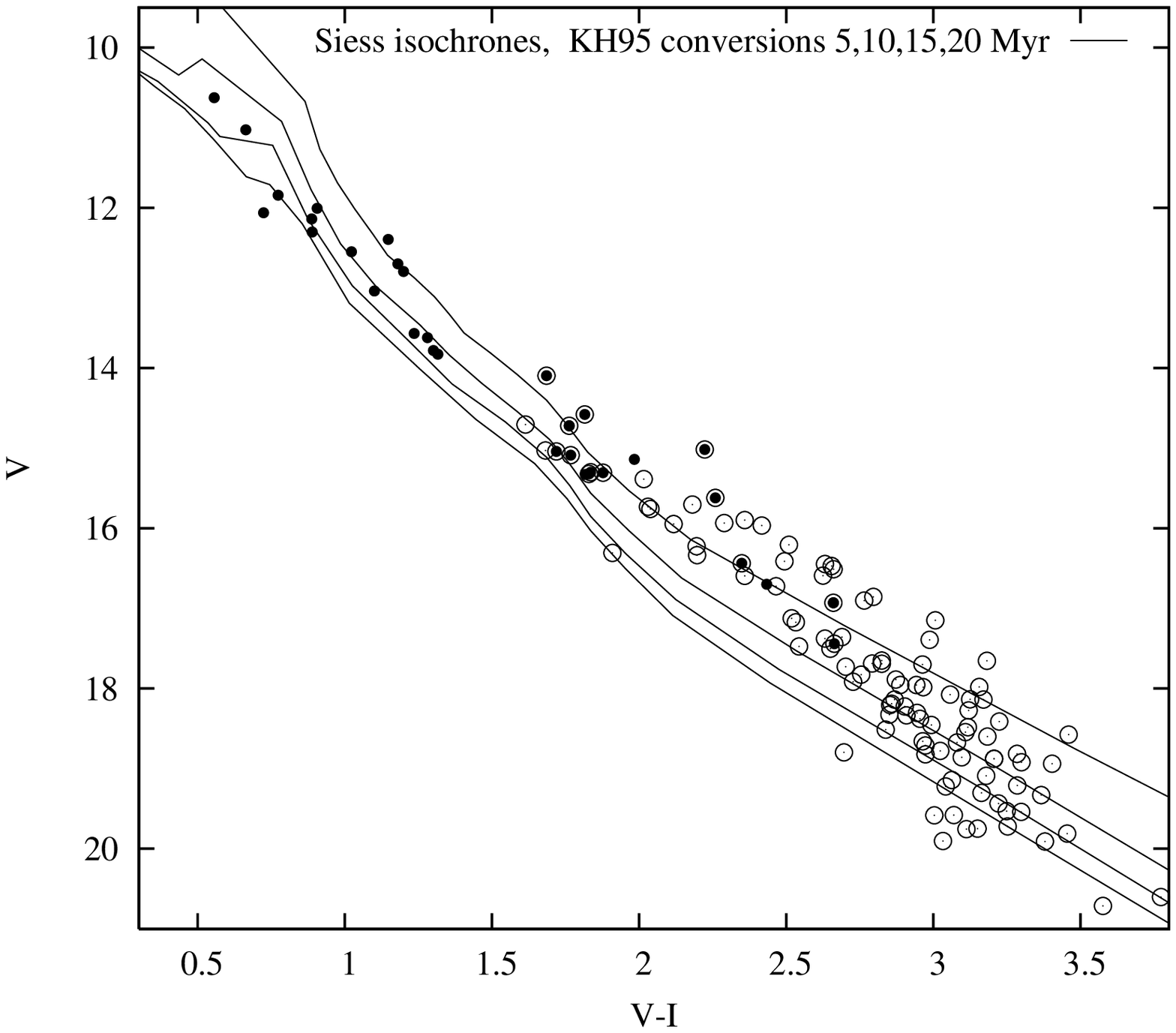}
\end{minipage}
\caption{(a) A comparison of low-mass $\gamma$ Vel PMS association
  members with evolutionary isochrones in the $V$ vs $V-I$ CMD. The
  isochrones have been shifted to an intrinsic distance modulus of
  7.76 mag, an extinction $A_V = 0.131$ and reddening $E(V-I)=0.055$.
  The isochrones are calculated from the $Z=0.02$ Siess et al. (2000)
  models, using a colour-temperature relationship tuned to match the
  Pleiades (see Naylor et al. 2002). Dots show spectroscopically
  confirmed members, open circles are X-ray selected members. 
  (b) A similar comparison using the Baraffe et al. (2002) models with
  a mixing length of 1.0 pressure scale heights. (c) A similar comparison using
  the D'Antona \& Mazzitelli (1997) models. (d) A similar comparison to
  (a) but using the colour-temperature conversions from Kenyon \&
  Hartmann (1995).}
\label{pmsfit}
\end{figure*}

\begin{figure*}
\centering
\begin{minipage}[t]{0.45\textwidth}
\includegraphics[width=80mm]{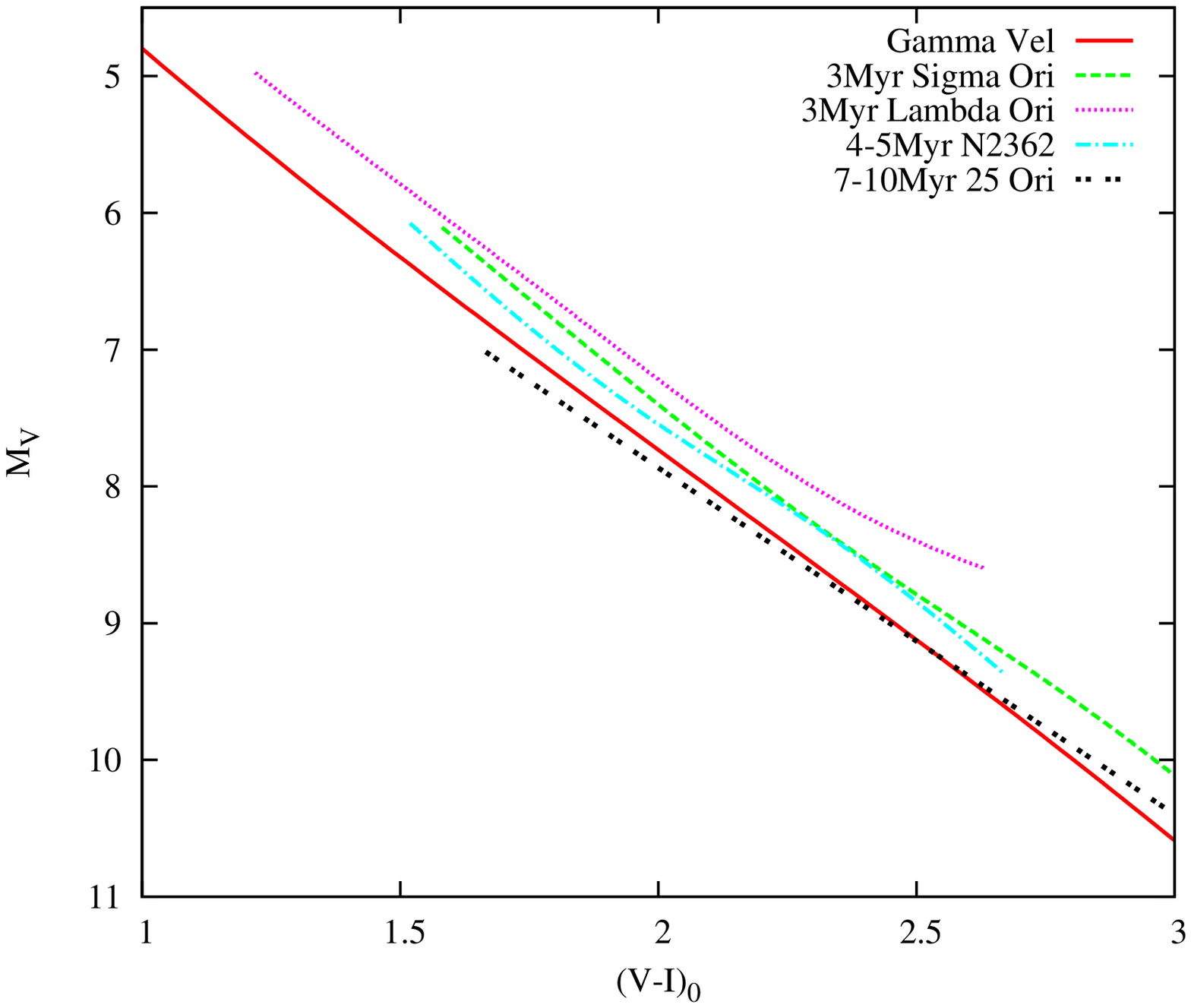}
\end{minipage}
\begin{minipage}[t]{0.45\textwidth}
\includegraphics[width=80mm]{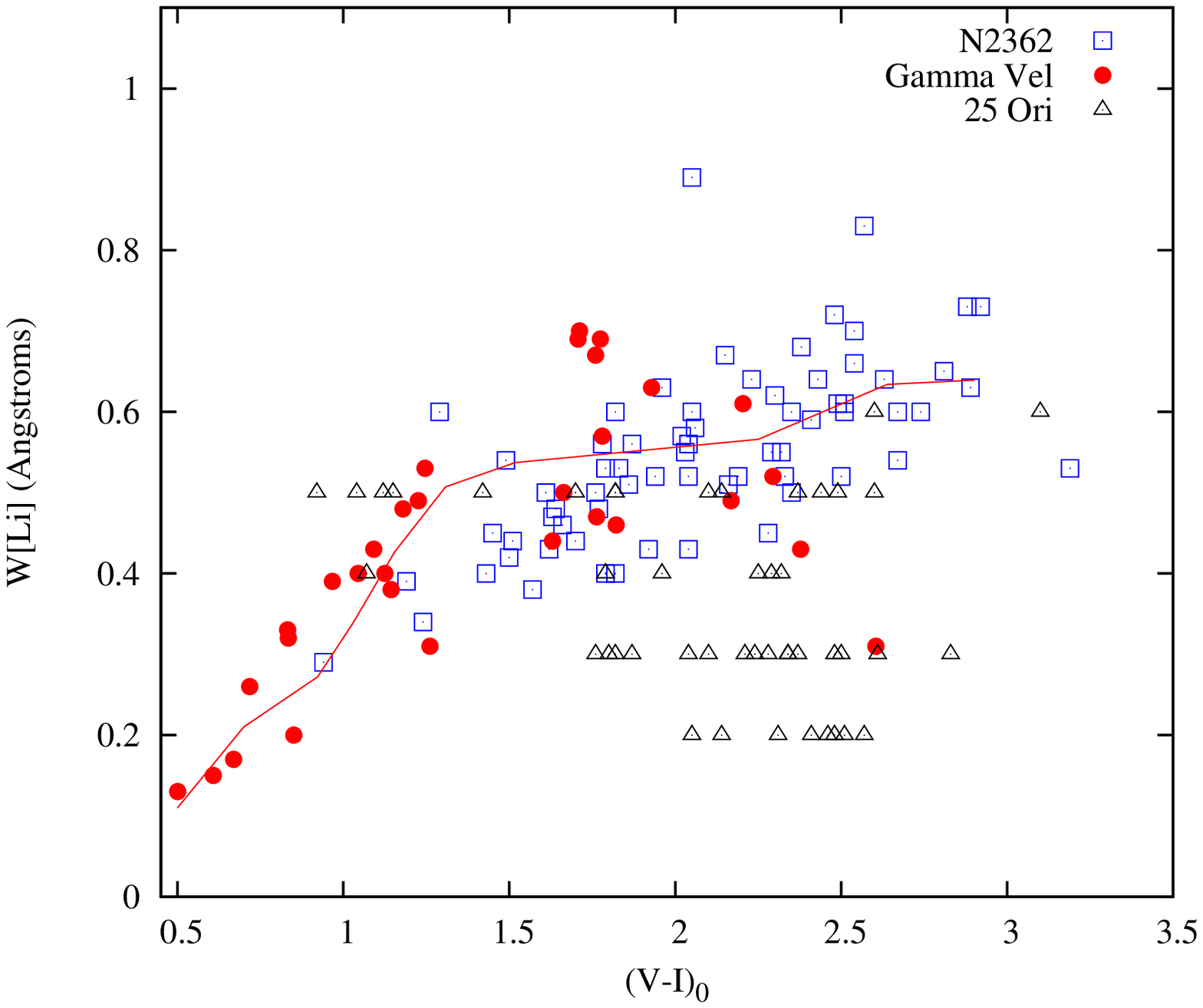}
\end{minipage}
\caption{
(a) A comparison of the $\gamma$ Vel PMS association empirical
  isochrone with empirically determined isochrones for other
  clusters/associations in the absolute $V$ vs intrinsic $V-I$
  CMD. These isochrones have been shifted according to their individual
  intrinsic distance modulus, extinction and reddening (see text).  The
  red solid line is a third order polynomial fit to the $\gamma$ Vel
  PMS data compared to similar representations for the $\sigma$ and
  $\lambda$ Ori clusters (age 3\,Myr, Mayne \& Naylor 2008), NGC 2362
  (age 4--5\,Myr, Mayne \& Naylor 2008) and the 25~Ori cluster (age
  7--10\,Myr, Brice\~no et al. 2007). (b) An empirical comparison based
  on the EW of the Li~6708\AA\ feature versus intrinsic $V-I$ colour.
  Data are shown for NGC~2362, the 25~Ori cluster and the PMS around
  $\gamma$~Vel.  The W[Li] values for 25~Ori are quantised in 0.1\AA\
  steps.  The solid line shows the locus of undepleted Li defined in
  section~\ref{specmembers}.
}
\label{nathan}
\end{figure*}

A model dependent absolute age can be estimated by comparing the PMS
data to model isochrones.  In practice this is difficult to do. Figure
\ref{pmsfit} shows various attempts to match a PMS sample around Gamma
Vel, consisting of the spectroscopic and X-ray selected photometric
members defined in section~\ref{members}, using different models.  The
top left panel shows Siess et al (2000) isochrones overlaid on the
data.  These are the $Z=0.02$ models, with a relationship between colour
and effective temperature that is tuned to match photometry in the
Pleiades, as described in Naylor et al (2002). If we restrict ourselves
to $V-I<2.5$ where we know our photometric calibration is excellent,
then the lower envelope of the data suggests ages of 10--12\,Myr, whilst
the upper envelope suggests ages $<5$\,Myr. Of course our data
doubtless contain many binaries which are up to 0.75 mag brighter than
a single star of the same colour, but even so, the redder stars have a
magnitude spread which is larger than this, perhaps indicating a real
age spread.  For $V-I>2.5$ the data dip well below the 10\,Myr
isochrone in a way not reproduced by the Siess et al isochrones,
either due to inadequacies in the cool model atmospheres or possibly in
the photometric calibrations for very red stars.

The top right panel of Fig.~\ref{pmsfit} shows a similar comparison
using the mixing length 1.0 isochrones from Baraffe et al. (2002),
again with the bolometric corrections and colour-effective-temperature
relations outlined in Naylor et al (2002). The bottom left panel shows
the same comparison with the models of D'Antona \& Mazzitelli
(1997). The data for $V-I<2.5$ in these diagrams suggest that the
oldest single stars have ages of 15--20\,Myr and 7--10\,Myr
respectively, with the same discrepancies for cooler stars. The
distance uncertainty for the $\gamma$ Vel association makes little
difference to these numbers and the reddening vector runs almost
parallel to the isochrones.

A further factor to be considered is the choice of colour-temperature
conversion relation. In the bottom right panel of Fig.~\ref{pmsfit} we
show the results of transforming the Siess isochrones using the
often-used colour-temperature relationship of Kenyon \& Hartmann
(1995). Obviously the data are far from parallel to these
isochrones. The age of the association would be 5--15\,Myr, depending
on which colour range was considered. Finally, all these models are
roughly at a solar metallicity. Relaxation of this assumption is
discussed in section~\ref{discussion}.

\subsubsection{Low-mass PMS empirical isochrones}

An alternative approach is to consider empirical isochrones.  In
Fig.~\ref{nathan}a we compare our data to empirical isochrones for the
$\sigma$ Ori, $\lambda$ Ori, NGC~2362 and 25~Ori clusters. These
isochrones were constructed by fitting low-order polynomials or splines
through published datasets of low-mass members (see Mayne et al. 2007;
Mayne \& Naylor 2008) and adjusting the distance moduli and reddening
to match those assumed for the $\gamma$~Vel association. For $\sigma$
Ori, $\lambda$ Ori and NGC~2362 we used distances and reddening
consistently rederived for early-type members (using the same
techniques as described here) by Mayne \& Naylor (2008), whereas for 25
Ori we used the intrinsic distance modulus of 7.59 and $A_V=0.29$
derived by Brice\~no et al. (2007).

Figure \ref{nathan}a suggests that the PMS stars around $\gamma$ Vel
are older than those of the $\sigma$ and $\lambda$ Ori
clusters, which have been assigned ages of $\simeq 3$\,Myr in the
empirical age ladder of Mayne \& Naylor (2008), older than NGC~2362
which is assigned an age of 4--5\,Myr and similar in empirical age to
the PMS stars of the 25~Ori cluster, which Brice\~no et al. (2007)
concluded had an age of 7--10\,Myr, based mainly on Siess isochrones. A
similar empirical analysis for the $\gamma$~Vel association 
using $V-J$ colours was performed by
Hern\'andez et al. (2008, see their Fig.~9), assuming an intrinsic
distance modulus of 7.72 and $A_V=0.15$. 
They also concluded that the $\gamma$ Vel PMS stars were
older than those in $\sigma$ and $\lambda$ Ori, but found they were
younger than those in 25~Ori.

\subsubsection{Lithium depletion}

Lithium is strongly depleted in cool stars as they age. The W[Li]
values for the $\gamma^2$~Vel PMS candidates are shown in
Fig.~\ref{lirv}b, along with a locus marking the level of undepleted
Li.  There is no strong evidence for Li depletion in any of the
candidates with the appropriate RV (see
section~\ref{specmembers}).  Theoretically, as a cluster ages we expect
to see Li depletion beginning in PMS stars with effective temperatures
of 3600--3800\,K at ages of 10--20\,Myr (Baraffe et al. 1998; Siess et
al. 2000, see also Fig.~9 of Jeffries et al. 2003). At older ages a
``Li-depletion chasm'' opens up towards both higher and lower
temperatures. A $T_{\rm eff}$ of 3600--3800\,K corresponds to an
observed $V-I$ of 2.65--2.25 in this association, so we have observed a
few stars which might be expected to show Li-depletion if they were
older than 10\,Myr. That stars with $V-I>2.25$ show little or
no Li-depletion allows us to put an upper limit to their age of 20\,Myr
using the Baraffe et al. (1998) models (with mixing length parameter of
1.0 or 1.9 pressure scale heights) or 10\,Myr for the $Z=0.02$
Siess et al. (2000) models. The D'Antona \& Mazzitelli (1997) models
predict that all stars within $3600<T_{\rm eff}<4500$\,K
($2.6>V-I>1.3$) would be Li-depleted even at 5\,Myr, which is
clearly incompatible with the data.

We can also consider Li-depletion empirically. The comparison clusters
in Fig.~\ref{nathan}a also have W[Li] measurements, some of which we
show in Fig~\ref{nathan}b, a plot of W[Li] versus intrinsic colour. For
NGC~2362 we take values for photometry and W[Li] from Dahm
(2005). $E(V-I)$ was taken to be 0.04 mag following Mayne \& Naylor
(2008). For 25 Ori the photometry and W[Li] come from Brice\~no et
al. (2007) and we use their $E(V-I)$ of 0.12 mag. Differences in the
definition of continuum levels and spectral resolution 
in the various analyses may lead to
systematic offsets in the W[Li] values, but it is clear that the
PMS stars around $\gamma$ Vel are more akin to those of NGC~2362 in
terms of Li depletion.  In 25~Ori there seems to be a population of Li-depleted
stars that are not present in the $\gamma$~Vel sample.  If it
is assumed that 25~Ori has an age of 7--10\,Myr then the PMS around
$\gamma$~Vel appears to have an age less than this. A caveat is that
the 25~Ori data have much lower resolution ($\simeq 6$\AA) than the
spectra analysed here.

\subsubsection{Summary of PMS age estimates}

In summary, from isochrones in the CMD, PMS stars around $\gamma$ Vel
are at ages of 5--10\,Myr on the age scale commonly adopted for other
young star forming regions and clusters. There may be a small spread in
ages around a single value. This empirical age agrees most closely with
the age obtained by comparison with the D'Antona \& Mazzitelli (1997)
and Siess et al. (2000) isochrones, but is much younger than the age
indicated by the Baraffe et al. (2002) isochrones.  The evidence from
the (absence of) lithium depletion also gives an age $<10$\,Myr according to the
Siess et al. (2000) models and an empirical age of $<10$\,Myr by
comparison with other clusters. For the rest of the paper we adopt an
age of 7\,Myr, but ages of 5--10\,Myr are possible.

\subsection{The age and mass of $\gamma$~Vel}

\label{gamvelage}

North et al. (2007) use their interferometric distance to estimate an
absolute $V$ magnitude of $-5.63\pm0.10$ for the O-star component of
$\gamma^2$~Vel, and that it has a mass and age of
$28.5\pm1.1\,M_{\odot}$ and $3.5\pm 0.4$\,Myr using the
solar-metallicity evolutionary models of Meynet et al. (1994).  This
is significantly younger than our estimates for the
low-mass PMS population. Even incorporating rapid rotation
into the model (e.g. Meynet \& Maeder 2003) would not increase this age
by more than 20 per cent.

The initial mass and lifetime of the WC8 component are more
uncertain. North et al. (2007) find an absolute magnitude of $-4.23 \pm
0.17$ and a present day mass of $9.0\pm0.6\,M_{\odot}$. If we assume
that the eccentric orbit of $\gamma^2$~Vel makes it unlikely that there
has been Roche-lobe overflow during its lifetime, then the WC8
progenitor probably did not become a red supergiant, which leads to a
lower limit to its initial mass of $M>40\,M_{\odot}$ according to the
models of Meynet \& Maeder (2003).  The same models suggest that the
age of a (non-rotating) $M>40\,M_{\odot}$ WC8 star is $<5$\,Myr. 
Agreement with the O-star age of 3.5\,Myr would require the initial mass of
the WC8 component to be 60--80\,$M_{\odot}$.

$\gamma^1$~Vel does not feature in the
Hipparcos or Tycho catalogues, but has $V=4.27$, $B-V=-0.23$ (from the
preliminary 5th revised edition of the bright star catalogue by
D. Hoffleit and W. H. Warren [1991] held by the NASA Astronomical Data
Center), a spectral type of B2III and is an SB1 system (Hern\'andez
\& Sahade 1980). For the distance modulus and extinction derived for
the MS around $\gamma$~Vel this gives an absolute magnitude of -3.62
and an intrinsic $B-V$ of -0.27. If we assume uncertainties of $\pm
0.02$ on the photometry and compare with the solar metallicity evolutionary models of
Lejeune \& Schaerer (2001), the age could be anywhere between 3
and 15\,Myr, with a best fit at an age of 8\,Myr and a mass of
14\,$M_{\odot}$.

\subsection{Circumstellar Material}

\begin{figure*}
\centering
\begin{minipage}[t]{0.45\textwidth}
\includegraphics[width=80mm]{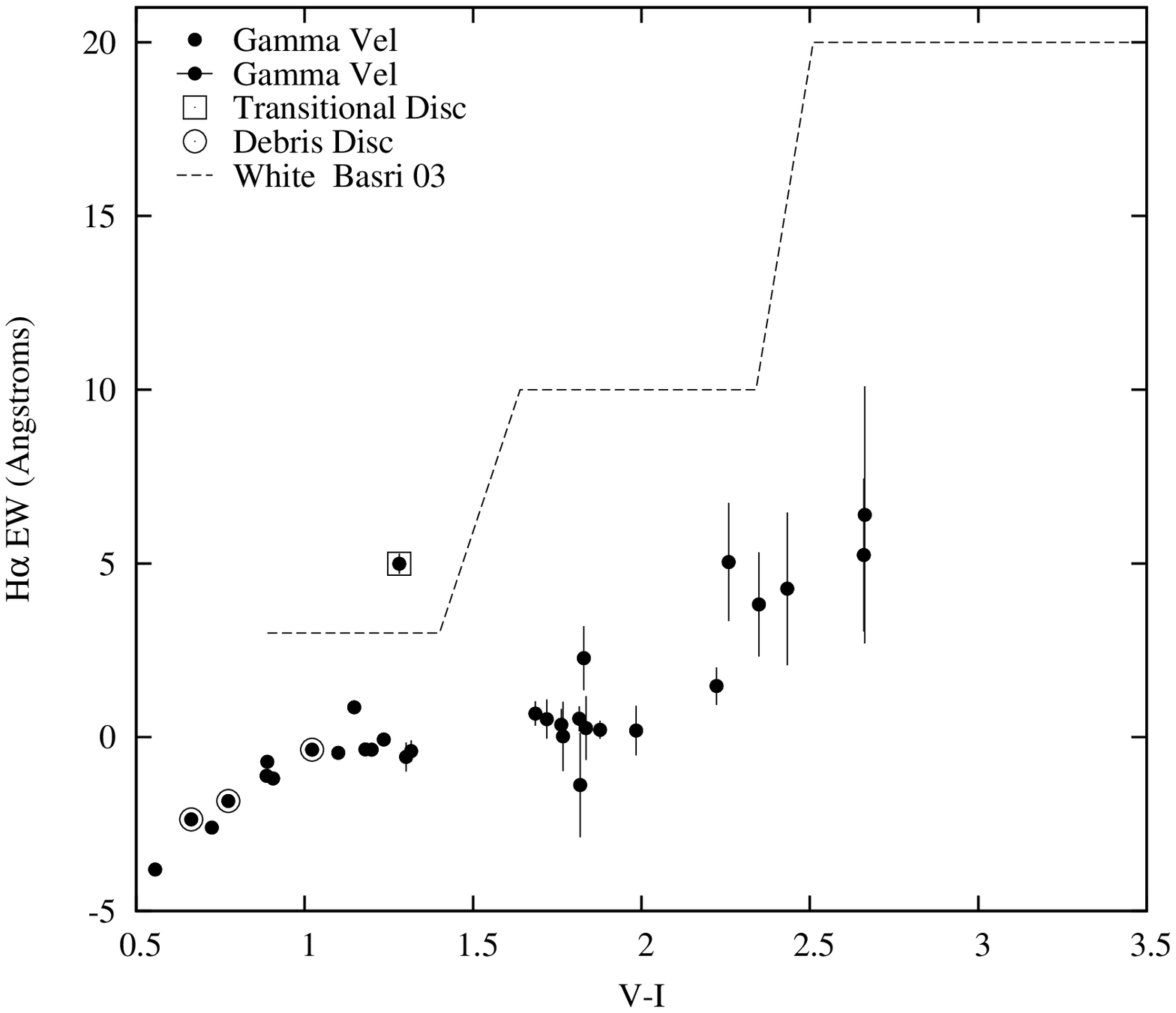}
\end{minipage}
\begin{minipage}[t]{0.45\textwidth}
\includegraphics[width=80mm]{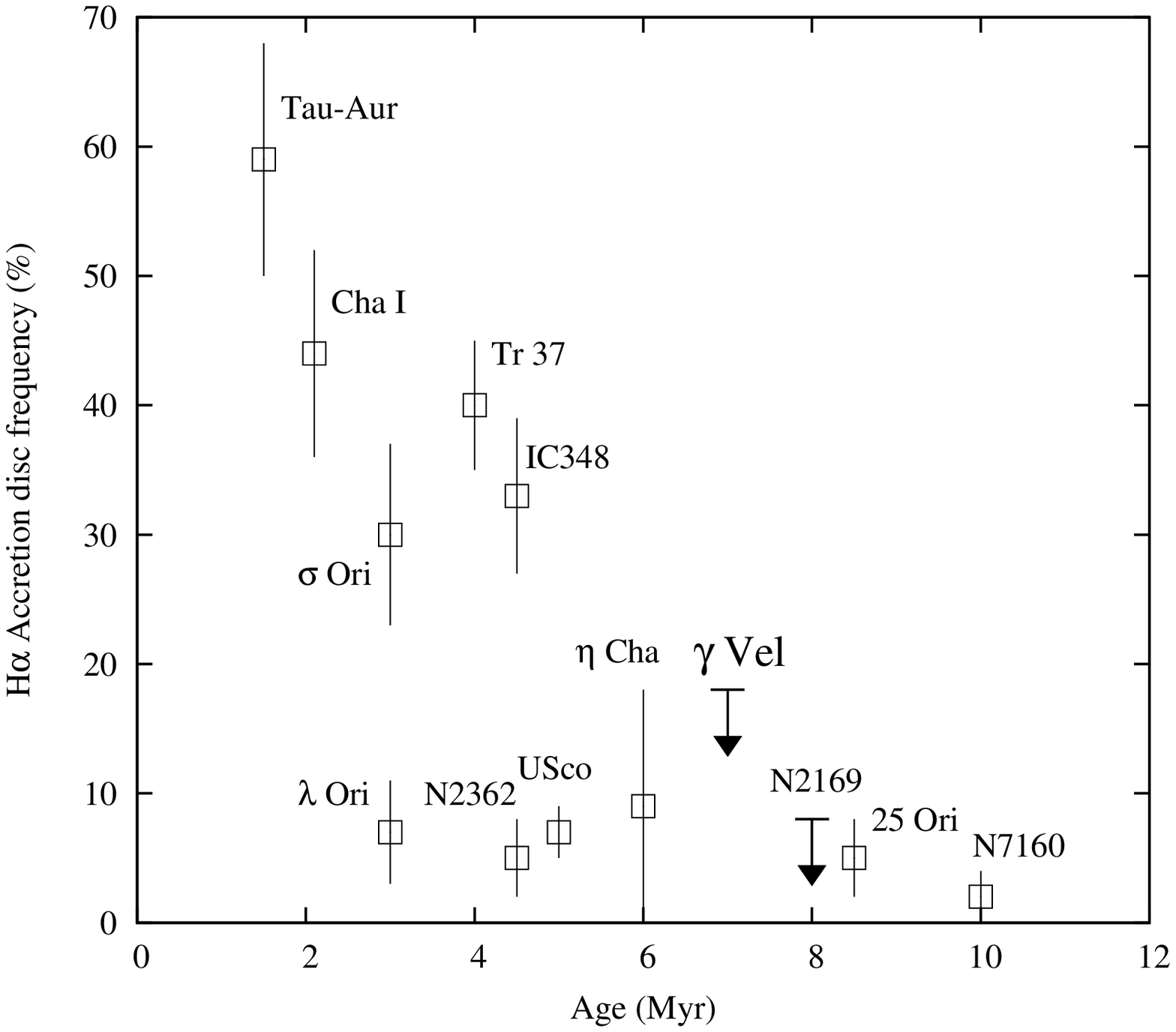}
\end{minipage}
\caption{
(a) The H$\alpha$ EWs of spectroscopically confirmed members of the
  $\gamma$~Vel association compared with an empirical locus (defined by
  White \& Basri 2003) that separates accreting CTTS from non-accreting
  WTTS. Also indicated are several stars with indications of
  circumstellar material in Spitzer near- and mid-IR observations by
  Hern\'andez et al. (2008).
  (b) The fraction of low-mass stars ($V-I>1.4$) in the $\gamma$~Vel
  association that have accretion signatures (according to the White \&
  Basri [2003] H$\alpha$ EW criteria) compared with K5-M5 stars in
  other clusters and associations as a function of age.  The data come
  from Mohanty, Jayawardhana \& Basri (2005) for Taurus-Auriga, IC~348,
  Chamaeleon I and Upper Sco, from Dahm (2005) for NGC 2362, from
  Sicilia-Aguilar et al. (2005) for Tr~37 and NGC~7160, from
  Jayawardhana et al. (2006) for the $\eta$~Cha group, Brice\~no et
  al. (2007) for 25~Ori, Jeffries et al. (2007) for NGC~2169 and from
  Sacco et al. (2008) for $\sigma$~Ori and $\lambda$~Ori. The relative
  precisions for the cluster ages are typically $\pm 1$--2\,Myr.
}
\label{accreteplot}
\end{figure*}

Young PMS stars often show H$\alpha$ emission, either as a consequence
of rapid rotation and magnetically generated chromospheric activity or
due to continued accretion from a circumstellar disk. The H$\alpha$
emission line in accreting stars (often called classical T-Tauri stars -
CTTS) is systematically stronger and broader than in weak lined T-Tauri
stars (WTTS), where the emission appears to be mainly chromospheric.

In Fig.~\ref{accreteplot}a we show the EW of the H$\alpha$ lines of the
spectroscopically confirmed $\gamma$~Vel members as function of
colour. Also shown is an empirical dividing line between CTTS and WTTS.
White \& Basri (2003) defined this dividing line in terms of spectral
type and it has been converted to observed $V-I$ colour using the
conversion table of Kenyon \& Hartmann (1995).

Hern\'andez et al. (2008) analysed Spitzer infrared observations of the
$\gamma$~Vel association members. On the basis of near- and
mid-infrared colours they were able to classify stars as discless
(class III), having optically thick inner dust discs (class II), discs
in transition between class II and class III (transitional or
pre-transitional discs) or stars showing only a mid-IR excess
(debris discs). In our sample of 32 spectroscopic members there are
three examples of stars with debris discs and one with a pre-transitional disc. 
The rest are classed as discless class III objects. The 
stars with discs are indicated in Fig.~\ref{accreteplot}a.

From Fig.~\ref{accreteplot}a we see that only the star classified as a
pre-transitional disc object has an H$\alpha$ EW large enough to be
considered a signature of accretion (star ID 16-40, see
Fig.~\ref{specplot}). This is also the only H$\alpha$ profile with a
width greater than the that accretion threshold of 270\kms\ suggested
by White \& Basri (2003).  All the other objects have H$\alpha$ EWs
consistent with chromospheric emission alone. The stars with debris
discs do not show any anomalous behaviour. The low frequency of
accreting objects is consistent with the low frequency of class II disc
sources found by Hern\'andez et al. (2008).

The disappearance of accretion signatures as young clusters become
older is a probe of infalling gas and the evolution of inner discs.
Fig.~\ref{accreteplot}b shows the fraction of stars classified as CTTS
using the White \& Basri (2003) criteria as a function of age for a
number of young star forming regions and clusters. In all cases the
samples have been restricted to stars with spectral type between K5 and
M5 ($V-I>1.4$ in the $\gamma$~Vel association). 
The cluster ages are from Mayne \& Naylor (2008) where available or
we have chosen published ages based on the Siess et al. (2000)
isochrones -- which are the most consistent with Mayne \& Naylor's
empirical age scale.

The 95 per cent upper limit to the accretion disc frequency in the
$\gamma$~Vel (as diagnosed from the H$\alpha$ EW) is 18 per cent. There
are ample numbers of spectroscopically unobserved candidate members
with which to refine this value, but the present result is consistent
with the idea that, with rare exceptions, accretion-related H$\alpha$
emission is absent for stars older than 5\,Myr.

\subsection{The total mass and mass function of the Gamma Vel
  association}

\label{massfunction}

In section~\ref{photomembers} 319 stars were defined as low-mass
photometric PMS candidates with $V<20$ and $V-I>2$.  This sample is
almost complete ($\sim 85$ per cent of such stars have unflagged
photometry -- see section~\ref{photometry}) and there will be $<117$
contaminating field stars within this sample if the density of
contaminants is similar to our estimate within the {\it XMM-Newton}
field of view (see section~\ref{kinspace}).

If we assume a
distance modulus of 7.76 mag, an extinction of 0.131 mag and an age of
7\,Myr, the models of Siess et al. (2000) translate the boundaries of
this photometric selection to an approximate mass range of
$0.1<M/M_{\odot}<0.6$, with little change for reasonable variations
in these assumptions. Correcting for incompleteness, we can 
say there are between 238--375 stars in this mass range for the area
covered by our photometric survey. Poissonian uncertainties are negligible
compared with uncertainty in the contamination fraction.
If we assume the ``universal'' mass function advocated by Kroupa (2001)
($\psi(M) \propto M^{-2.3}$ for $M\geq 0.5\,M_{\odot}$ and $\psi(M)
\propto M^{-1.3}$ for $0.08<M/M_{\odot}<0.5$), then we expect to see
9--14 higher mass stars with $1.8<M/M_{\odot}<4.4$, corresponding to
$7<V_T<10$ at the distance/extinction of the association, according to the
solar metallicity models of Lejeune \& Schaerer (2001). Consulting
Fig.~\ref{pm_sel}, there are actually 22 stars on the main sequence in
this range, 14 of which are within our photometric survey
area, though we cannot be absolutely sure that all of these are
association members. Hence the observed ratio of high- to low-mass
stars is consistent with the assumed mass function.

We can go a step further and ask how likely it is to find stars as
massive as $\gamma^1$~Vel and $\gamma^2$ Vel in this association?
$\gamma^1$~Vel is 14\,$M_{\odot}$ and the O-star component of
$\gamma^2$~Vel is 28.5\,$M_{\odot}$. The WC component is presently less
massive than the O-star, but was $>40\,M_{\odot}$ in the past (see
section~\ref{gamvelage}).  The number of stars more massive than
10\,$M_{\odot}$ predicted by the Kroupa mass function is 1.4--2.1,
whilst the number with $M>40\,M_{\odot}$ is expected to be only
0.1--0.2. Of course these predictions assume that all the association
low-mass stars have been found, but we know (see
section~\ref{kinspace}) that our survey has not yet reached the spatial
boundary of this young aggregate, so these predicted numbers are lower
limits. We conclude that although the relative numbers of low- and
high-mass stars is as expected for a Kroupa (2001) mass function, the
presence of a system as massive as $\gamma^2$ Vel may be unexpected.

The total mass of the association within the photometrically surveyed
area can be estimated by integrating the Kroupa mass function (normalised
to the low-mass candidates) up to 10\,$M_{\odot}$ and then adding the
masses of $\gamma^1$ and $\gamma^2$ Vel (for which membership is
certain by definition!). We apply an approximate correction factor of
1.25 to the integrated mass function to account for the likelihood that
about half of the stars in our mass function will be in binary systems
with a roughly flat mass-ratio distribution (see Jeffries et al. 2004).
The total mass is is 250--360\,$M_{\odot}$ depending on how much
contamination is present among the low-mass photometric candidates.  If
the additional high-mass stars found within a 45 arc-minute radius, but
outside our photometrically surveyed area are also association members,
and stars of different masses are spatially distributed in the same
way, then the total association mass could rise to
360--530\,$M_{\odot}$.  Of course at present there is no way of knowing
where the cluster around $\gamma$ Vel ends and the Vela OB2 association
begins, so a better question might be what is the total mass of the
Vela OB2 association if a Kroupa-like mass function is widely
applicable?

There are 74 objects selected as Vela OB2 members by de Zeeuw et
al. (1999), which lie close to the proposed main sequence and have
$4<V<9$, a range where the Hipparcos catalogue should be almost
complete. This magnitude range corresponds to $2.5<M/M_{\odot}<17$
according to the solar metallicity models of Lejeune \& Schaerer
(2001). Extrapolating to 0.1\,$M_{\odot}$ using the Kroupa mass
function, we would expect to find a further $\sim 2300$ association
members with $9<V<20$ and a total association mass of $\sim
1700\,M_{\odot}$.

\section{Discussion}

\label{discussion}

In this paper we have established that there is a large group of very
young, low-mass stars in the direction of $\gamma$~Vel which are both
spatially and kinematically coherent. The question is how are these
stars related to $\gamma$~Vel and the wider Vela OB2 association? This
relationship can be probed in terms of distance, kinematics and age.

In sections~\ref{fitting_ms} and~\ref{velaob2} we found that the
distance to the Vela OB2 association ($7.72\pm 0.08$) and the distance
to MS stars around $\gamma$ Vel ($7.76\pm 0.07$) agree with good
precision. These distances also fall between the two interferometric
distances to $\gamma^2$ Vel ($7.82^{+0.22}_{-0.07}$ -- Millour et
al. 2007; $7.63\pm 0.05$ -- North et al. 2007). In
section~\ref{massfunction} we showed that the numbers of low- and
high-mass stars around $\gamma$~Vel are in a ratio that is consistent
with a typical cluster or field star mass function.

The simplest explanations for these observations are that the MS stars
around $\gamma$~Vel are the high mass population commensurate with the
low-mass PMS stars we have discovered, that $\gamma^2$ Vel is at a
similar distance to these stars and that they are all part of the Vela
OB2 association.  Support for this hypothesis is provided by the
significant concentration of low-mass stars around $\gamma$~Vel (see
Fig.~\ref{spatial}) and the very similar proper motions of
$\gamma^2$~Vel, the MS stars around $\gamma$~Vel, the spectroscopically
confirmed low-mass association members and the Hipparcos-selected Vela
OB2 members (see ~\ref{kinspace}).

There are two potential problems with this model. Firstly, that a very
massive object like $\gamma^2$~Vel is unlikely to have formed in a
cluster consisting of only the stars we have currently identified as
$\gamma$~Vel association members. Weidner \& Kroupa (2006) present a
simulation (their Fig.~11) aimed at predicting the mass of the most
massive star in a cluster of total mass $355\,M_{\odot}$, which is
similar to the mass of the $\gamma$~Vel association within the area of
our photometric survey (see section~\ref{massfunction}).  For their
favoured ``sorted-sampling'' scenario, the most probable most massive
star is $M\simeq 15\,M_{\odot}$, with only a few per cent chance of
producing a star as large as the $\geq 40\,M_{\odot}$ initial mass of
the WC8 star in $\gamma^2$ Vel. Secondly, the age of $\gamma^2$~Vel is
well constrained from its evolutionary status and seems to be
significantly less (3--4\,Myr) than the age of the low-mass PMS stars
(5--10\,Myr), although these ages are dependent on the accuracy of
evolutionary models in quite different mass ranges.

Rather than problems, these two issues may be important clues to the
formation and history of the $\gamma$~Vel association and its
relationship with Vela OB2. Weidner \& Kroupa (2006) propose a scenario
whereby clusters form in an ordered fashion: the low-mass stars form
first and as molecular cloud contraction proceeds, larger amplitude
density fluctuations are generated which then form massive stars. This
ordering may be enhanced by the build up of the most massive stars
through competitive accretion or mergers (Bonnell \& Bate 2005, 2006).
Once one or more massive stars form near the centre of a cluster then
radiative feedback and winds can become important, with the potential to deposit
significantly more energy into the intracluster medium than the
gravitational binding energy. This could drive out the remaining gas and
unbind the cluster (if it were bound to begin with). That the most
massive stars in a cluster form after the bulk of low-mass star
formation is complete has also been proposed to simultaneously explain why
most low-mass stars in the Orion Nebula cluster possess optically thick
inner discs, yet there are many ``proplyd'' objects close to the
central massive Trapezium stars with photoevaporation timescales as
short as 0.1\,Myr (Smith et al. 2005; Clarke 2007)

A possible scenario for the formation of the $\gamma$~Vel association
is as follows. Star formation began about 5--7\,Myr ago, with a cluster
of low-mass stars forming first. After 1--2\,Myr the $\gamma^2$~Vel
system formed and the Lyman continuum flux from the $\geq
40\,M_{\odot}$ WC8 progenitor and the O-star component
($\leq$O5V$+$O7.5V, $\geq 3\times 10^{49}$ ionising photons\,s$^{-1}$
-- Martins, Schaerer \& Hillier 2005) could rapidly ionise and drive
out the remaining gas, terminating star formation in its vicinity.  At
this stage most of the cluster mass is in the form of gas, so its
expulsion is likely to unbind the cluster (e.g. Weidner et al. 2007).
The present kinematic data offer only an upper limit of
$\sim$2\,km\,s$^{-1}$ on the 1-dimensional velocity dispersion of
association members, but the escape velocity for stars in our
photometric survey is only $\simeq 0.5$\,km\,s$^{-1}$.  Unbound
association members would then drift for about 3\,Myr and during this
time the association could expand from an initially compact
configuration to a diameter of 6--12\,pc, corresponding to 1--2 degrees
on the sky. It is quite possible that much of the initial stellar
population of the association now lies outside the 1 degree field of
our photometric survey (as suggested in section~\ref{kinspace}) and
that the total initial stellar mass could be closer to the $\sim
1000\,M_{\odot}$ that Weidner \& Kroupa (2006) suggest should be
associated with the formation of a $40\,M_{\odot}$ star. A wider
photometric survey could confirm this suggestion and may find the
``edge'' of the distribution of low-mass stars associated with
$\gamma$~Vel.

The scenario described above does not account for the rest of Vela OB2,
which evidences recent star formation over a diameter that is 5 times
larger. Most of these other early-type stars cannot have been in a more
compact configuration within the $\gamma$~Vel association, because
there has been insufficient time for them to travel to their present
positions. In section~\ref{massfunction} we found that the $\gamma$~Vel
association we have observed contains 15--20 per cent of the
anticipated low-mass stars (and total mass) of the Vela OB2 association
(providing the mass function is spatially constant), but this fraction
is found within only $\sim 1$ per cent of the area covered by Vela
OB2. This indeed suggests that the $\gamma$ Vel association is a
subcluster within the Vela OB2 assocation and is supported by the
clustering of the low-mass stars around $\gamma$~Vel shown in
Fig.~\ref{spatial}.

On the other hand, the coherence of their proper motions
suggests that all these stars did at least form in the same molecular
cloud. There are at least two possible explanations: (1) Star formation
was triggered across the cloud, perhaps by some external event like a
nearby supernova (in the older, but nearby Trumpler\,10 cluster for
example -- de Zeeuw et al. 1999). Star formation then proceeds as
described by the hydrodynamic simulations of Clark et al. (2005), with
the development of a number of spatially distinct, close-to-coeval
subclusters, which become unbound and disperse as they form more
massive stars. (2) Star formation could progress in a sequential
manner, with the expanding H\,{\sc ii} region of an initial generation
of high-mass stars compressing the surrounding molecular material,
triggering new bursts of star formation (e.g. Dale, Bonnell \&
Whitworth 2007).

A very wide photometric survey of the Vela OB2 association should be
capable of distinguishing between these scenarios, using low-mass PMS
stars as age probes. In the first model, the PMS stars will show little
or no spatial concentration around the early-type stars (at least no
more than we have found around $\gamma$~Vel) and should have only
small, spatially incoherent age variations.  In the second model we
would expect to find distinct age gradients (becoming younger radially
outwards from $\gamma$~Vel if it marks where star formation commenced),
with younger PMS stars being more closely clustered around their
high-mass siblings.

Throughout this paper we have assumed that
$\gamma$~Vel and its association have a roughly solar metallicity.  It
is worth investigating (prompted by the referee) what might happen to
the scenario discussed above if this assumption is relaxed. (1) If we
assume that the metallicity is roughly half-solar, the main-sequence
fitting distance moduli would decrease by $\sim 0.4$~mag, but the
estimated reddening would remain approximately unchanged (Mayne \&
Naylor 2008). This would place the main-sequence stars around
$\gamma$~Vel and the stars of Vela OB2 in front of the interferometric
distances to $\gamma$~Vel by $\sim 0.3$\,mag, casting doubt on
the association of $\gamma$~Vel with the surrounding main-sequence and
PMS stars. A twice-solar metallicity would increase the
main-sequence fitting distances leading to better agreement with the
mean Hipparcos parallax of Vela OB2 (de Zeeuw et al. 1999)
but even worse agreement with the distance to $\gamma$~Vel.  (2) The
$Z=0.01$ low-mass isochrones of Siess et al. (2000) are intrinsically about
0.25~mag fainter than the Z=0.02 isochrones. However, taking into
account the $\sim 0.4$ mag decrease in the distance modulus we would get
a slightly older age from modelling the PMS in the CMD. The empirical
cluster comparison would only be affected by the distance modulus
change and Fig.~15a shows that a 0.4~mag decrease in distance modulus
would lead to the $\gamma$ Vel PMS being significantly older than
25~Ori. A twice-solar metallicity would yield an age more similar to
Sigma/Lambda Ori.  (3) Meynet et al. (1994) and Meynet \& Maeder (2005)
consider the evolution of very high-mass stars of differing metallicity. A
half-solar metallicity would increase the deduced age and mass of the
O-star component of $\gamma$~Vel to 4.5\,Myr and 34\,$M_{\odot}$. A
half-solar metallicity progenitor to the WC8 component would have to be more
massive than 60\,$M_{\odot}$ to avoid a red supergiant phase and the
total stellar lifetime still less than 5\,Myr.

In summary a lower metallicity could increase the age of $\gamma$~Vel
but at the expense of also significantly increasing the deduced age for
the low-mass PMS.  A twice-solar metallicity could bring the ages of
the PMS and $\gamma$~Vel into agreement and also yield a distance for
the main-sequence around $\gamma~Vel$ that is similar to the Hipparcos
distance to Vela OB2. However in either the low or high-metallicity
scenarios there would be a very significant disagreement between the
interferometric distance to $\gamma^2$~Vel and the main sequence fit to
the bright stars around it. From the evidence provided by the spatial
and kinematic coherency of $\gamma$~Vel and its surrounding association
we do not consider either possibility likely.

\section{Summary}

The main findings of this paper can be summarised as follows:

\begin{itemize}
\item
We have extended the survey of Pozzo et al. (2000) and
photometrically identified a group of several hundred low-mass 
PMS stars surrounding the
high-mass spectroscopic binary system $\gamma^2$~Vel and its early-type
common proper-motion companion $\gamma^1$~Vel (which together we refer
to as the $\gamma$~Vel system). The youth of
a subsample of these objects has been qualitatively confirmed by the presence of
lithium in their atmospheres, H$\alpha$ emission and high levels of
X-ray activity.
\item
The spectroscopically confirmed PMS stars show coherence in their
radial velocities and proper motions. They share a common proper motion
with $\gamma$~Vel, a sample of high-mass stars in the vicinity of
$\gamma$~Vel and the Vela OB2 association. There is evidence for a
spatial concentration of the PMS stars around $\gamma$~Vel, but also
evidence that the ``edge'' of this association has not been reached in
our 0.9 square degree survey.
\item The number of higher mass main-sequence stars surrounding $\gamma$~Vel is
  commensurate with the number of lower mass PMS stars in the region
  according to the Kroupa (2001) ``universal'' mass function. The total mass of
  the association within our surveyed area is 250--360\,$M_{\odot}$.
  The intrinsic distance
  modulus of the higher mass stars, determined by main-sequence fitting, 
  is $7.76\pm 0.07$\,mag. This
  agrees well with a similarly-determined  distance modulus of $7.72\pm
  0.08$\,mag for the
  Vela OB2 association as a whole and is also in accord with interferometric
  distance determinations to $\gamma^2$~Vel. There now seems little
  doubt that the low-mass PMS stars are associated with $\gamma$~Vel
  and that these are all part of the Vela OB2 association.
\item The age of $\gamma^2$~Vel is well constrained by current high-mass
  stellar models to be less than 5\,Myr and more likely 3--4\,Myr.
  Fitting low-mass isochrones to the PMS population suggests much older
  ages ($\simeq 10$\,Myr) and empirically placing the PMS stars in an
  age sequence with other well-studied young clusters on the basis of
  their position in the Hertzsprung-Russell diagram and lithium depletion
  suggests ages of 5--10\,Myr.
\item Accretion activity, as judged by the levels of H$\alpha$ emission
  has almost entirely ceased in the PMS stars. The only example of an
  accretor when defined in this manner has been previously identified
  as possessing a ``transitional'' disc on the basis of its infra-red colours.
\item We can speculate that $\gamma^2$~Vel was formed after the bulk of
  the low-mass population and its ionising radiation may have been
  responsible for driving out gas, terminating star formation and
  unbinding the ``$\gamma$~Vel association''. The low observed velocity
  dispersion of the PMS population suggests that the whole Vela OB2
  association cannot have originated close to $\gamma^2$~Vel. The
  concentration of stars and stellar mass close to $\gamma$~Vel also
  suggests that it is a subcluster within Vela OB2. Star formation in
  Vela OB2 must have begun at several sites within the same molecular
  cloud, either sequentially, resulting in measurable age gradients or
  perhaps as a result of an external trigger, in which case PMS
  populations across Vela OB2 may be coeval.
\end{itemize}

\section*{Acknowledgments}
This work is based upon optical observations collected at the Blanco and 0.9-m
telescopes of the Cerro Tololo Interamerican Observatory and X-ray
observations collected by the {\it XMM-Newton} satellite and analysed
by the {\it XMM-Newton} Serendipitous Source Catalogue consortium led by the
University of Leicester.

Sadly, Christina (Tina) Devey died before the completion of this work
and this paper is dedicated to her memory.

We thank the referee (Sofia Randich) for a very careful examination of
the manuscript.

\nocite{meynet05}
\nocite{mohanty05}
\nocite{sicilia05}
\nocite{jayawardhana06}
\nocite{jeffriesn216907}
\nocite{dahm05}
\nocite{lejeune01}
\nocite{briceno07}
\nocite{mayne07}
\nocite{mayne08}
\nocite{meynet03}
\nocite{kenyon95}
\nocite{scelsi07}
\nocite{jeffries04}
\nocite{burningham03}
\nocite{naylor02}
\nocite{landolt92}
\nocite{pozzo00}
\nocite{hernandez80}
\nocite{millour07}
\nocite{north07}
\nocite{vanleeuwen07}
\nocite{dezeeuw99}
\nocite{schaerer97}
\nocite{smith68}
\nocite{dantona97}
\nocite{siess00}
\nocite{hog00}
\nocite{cutri03}
\nocite{naylor98}
\nocite{jeffries03}
\nocite{randich97}
\nocite{randich01}
\nocite{bohlin78}
\nocite{struder01}
\nocite{turner01}
\nocite{jeffries06lireview}
\nocite{schmutzrv97}
\nocite{raymond77}
\nocite{bohlin78}
\nocite{baraffe02}
\nocite{baraffe98}
\nocite{smith05}
\nocite{clarke07}
\nocite{clark05}
\nocite{dale07}
\nocite{weidner06}
\nocite{bonnell05}
\nocite{bonnell06}
\nocite{hernandez08}
\nocite{meynet94}
\nocite{martins05}
\nocite{kroupa01}
\nocite{zacharias05}
\nocite{bessell00}
\nocite{bessell98}
\nocite{pinsonneault98}
\nocite{weidner07}
\nocite{white03}
\nocite{sacco08}
\nocite{naylor06}

\bibliographystyle{mn2e}  
\bibliography{iau_journals,master}


\bsp 

\label{lastpage}

\end{document}